\documentstyle[bbm,amsfonts,epsfig,12pt,here]{article}

\newcommand {\eqref} [1] {(\ref {#1})}
\newcommand {\slsh} [1] {\not{\hbox{\kern-2pt${#1}$}}}

\newcommand {\beq} {\begin{equation}}
\newcommand {\eeq} {\end{equation}}
  \newcommand {\ber}{\begin{eqnarray*}}
  \newcommand {\eer} {\end{eqnarray*}}
\newcommand {\beqn}{\begin{eqnarray}}
  \newcommand {\eeqn} {\end{eqnarray}}

\newcommand{\Dslash}{\,{\raise.15ex\hbox{/}\mkern-12mu D}}
\newcommand{\Tr}{{\rm Tr}\,}
\newcommand{\gsim}{\lower.7ex\hbox{$
\;\stackrel{\textstyle>}{\sim}\;$}}
\newcommand{\lsim}{\lower.7ex\hbox{$
\;\stackrel{\textstyle<}{\sim}\;$}}

\def\href#1#2{#2}	

\def\N{{\cal N}}

\def\S_1{{\widetilde {S_1}}}

\def\R{{\mathbb R}}

\def\tr{{\rm Tr}}

\def\Dslash{{\rlap{\raise 1pt \hbox{$\>/$}}D}}
\def\O{{\cal O}}

\begin{document}
\begin{titlepage}
\begin{flushright}{UMN-TH-2632/08\,,  \,\,\, FTPI-MINN-08/03\,,  \,\, SLAC-PUB-13120}
\end{flushright}
\vskip 0.5cm

\centerline{{\Large \bf   QCD-like   
Theories on \boldmath{$R_3\times S_1$}: a Smooth }}

\vspace{2mm}

\centerline{{\Large \bf   Journey
from Small to Large \boldmath$r(S_1)$}}

\vspace{2mm}

\centerline{{\Large \bf 
 with Double-Trace Deformations}}

\vskip 1cm
\centerline{\large  M. Shifman${}^{a}$  and Mithat \"{U}nsal ${}^{b,c}$}

\vskip 0.3cm

\centerline{${}^a$   \it William I. Fine Theoretical Physics Institute,}
\centerline{\it University of Minnesota, Minneapolis, MN 55455, USA}

\vskip 0.2cm
\centerline{${}^b$ \it SLAC, Stanford University, Menlo Park, CA 94025, USA}
\vskip 0.1cm
\centerline{${}^c$ \it  Physics Department, Stanford University, Stanford, CA,94305, USA }

\vskip 1cm

\begin{abstract}

We consider QCD-like theories with one massless fermion in various
representations of the gauge group SU$(N)$. The theories are formulated on
$R_3\times S_1$. In the decompactification limit of large $r(S_1)$
all these theories are characterized by confinement, mass gap  and 
spontaneous breaking of a (discrete) chiral symmetry ($\chi$SB).  At small $r(S_1)$,
in order to stabilize the vacua of these theories at a center-symmetric point,
we suggest to perform a double trace deformation.
With these deformation, the theories at hand are at weak coupling at small
$r(S_1)$ and yet exhibit basic features of the large-$r(S_1)$ limit:
confinement and $\chi$SB. 
We calculate the string tension, mass gap, bifermion condensates and $\theta$ dependence.
The double-trace deformation becomes dynamically irrelevant
at large $r(S_1)$. Despite the fact that at small $r(S_1)$ confinement is Abelian,
while it is expected to be non-Abelian at large $r(S_1)$, 
we argue that small and large-$r(S_1)$ physics are continuously connected.
If so, one can use small-$r(S_1)$ laboratory to
extract lessons about QCD and QCD-like theories on $R_4$.

\end{abstract}

\end{titlepage}

\tableofcontents

\newpage

\section{Introduction}

Analyzing QCD and QCD-like theories on $R_3\times S_1$
 provides new insights in gauge dynamics at strong coupling
and offers  a  new framework for discussing various ideas on confinement.
The radius of the compact dimension $r(S_1)$ plays a role of an adjustable
parameter, an obvious bonus and a welcome addition to a rather scarce
theoretical toolkit available in strongly coupled gauge theories. 
As the circumference $L$ of the circle $S_1$ varies, so does the
dynamical pattern of the theory. For instance, at $L\ll \Lambda^{-1}$
in some instances  the theory becomes weakly coupled.
On the other hand, in the decompactification limit,  $L\gg \Lambda^{-1}$,
we recover conventional four-dimensional QCD, with its most salient feature,
non-Abelian confinement. 

A qualitative picture of confinement in terms of the Polyakov line
was suggested by Polyakov and Susskind  long ago \cite{Polyakov:1978vu, Susskind:1979up}.
Assume that the  compactified dimension is $z$.
The Polyakov line (sometimes called the Polyakov loop)
 is defined as a path-ordered holonomy of the Wilson line
in the compactified dimension,
\beq
{\cal U} = P\exp\left\{i\int_0^L a_z dz \right\} \equiv V U V^\dagger
\label{onem}
\eeq
where $L$ is the size of the compact dimension while
$V$ is a matrix diagonalizing ${\cal U}$,
\beq
U = {\rm diag}\{ v_1, v_2, ..., v_N\} \,.
\label{twom}
\eeq
According to Polyakov, non-Abelian confinement implies that the eigenvalues 
$v_i$ are randomized:  the phases of $v_i$ wildly fluctuate over the entire
interval $[0,2\pi]$ so that 
\beq
\langle {\rm Tr} U \rangle =0\,.
\label{threem}
\eeq
The exact vanishing of $\langle {\rm Tr} U\rangle$ 
in pure Yang--Mills 
is the consequence of the unbroken $Z_N$ center symmetry  in 
the non-Abelian confinement regime. Introduction of 
dynamical fermions (quarks)
generally speaking breaks the $Z_N$ center
symmetry at the Lagrangian level.\footnote{It is still an emergent dynamical symmetry 
in the multicolor limit \cite{Shifman:2007kt,Armoni:2007kd}; however, we limit ourselves to small $N$. In this paper parametrically $N$ is of order one.} However, the picture of wild fluctuations of the phases of $v_i$'s remains intact. Therefore, it is generally expected
that $\langle \frac{1}{N} {\rm Tr} U\rangle $ is  strongly suppressed even
with the   dynamical  fermion fields  that respect no  center symmetry,
$\langle \frac{1}{N} {\rm Tr} U\rangle \sim 0$. 
This expectation is supported by  lattice simulations at finite
temperatures \cite{L1} demonstrating 
that $\langle {\rm Tr} U\rangle$ is very close to zero at large $L$
(low temperatures).

\begin{figure}[t]
  \begin{center}
  \includegraphics[width=3in]{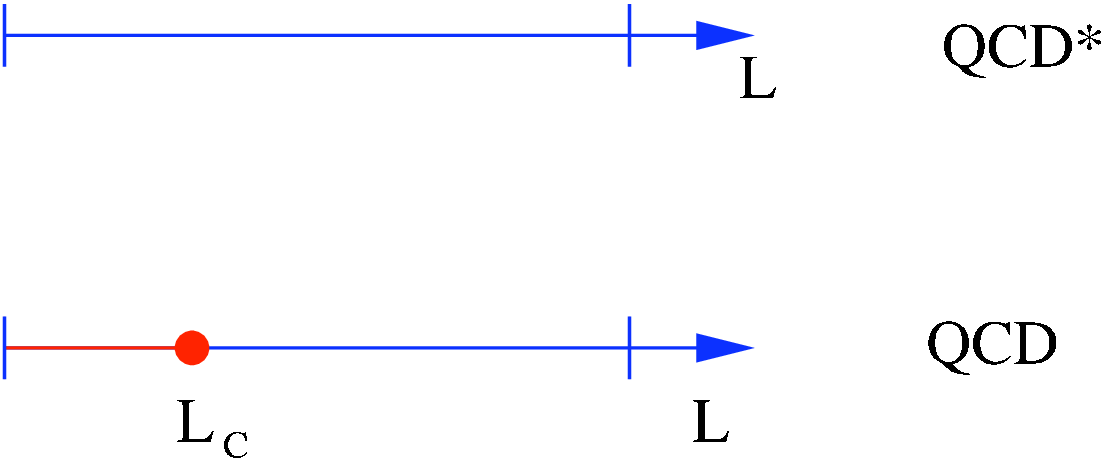}
  \caption
    {\small
Quantum chromodynamics as a function of compactified direction 
circumference before  and after surgery (QCD and QCD$^*$, respectively). 
$L_c$ is the point of a phase transition.}
   \end{center}
\label{fig:surgery}
\end{figure}

On the other hand, in QCD and QCD-like theories\,\footnote{
By QCD-like theories we mean non-Abelian gauge theories without elementary scalars,  
e.g., Yang--Mills with fermions in the two-index symmetric or antisymmetric representation, to be referred to as S/AS, see below.} at
small $L$ (high temperatures) the center-symmetric field configuration 
is dynamically disfavored. In many instances the vacuum is attained at
$\langle\frac{1}{N} {\rm Tr} U \rangle =1$. In this case, the effective low-energy theory is at strong coupling, and it is as hard to deal with it as with
QCD on $R_4$. Typically, the small and large-$L$ domains are separated
by a phase transition (or phase transitions). For instance, for S/AS
with even $N$ this is a $Z_2$ phase transition. 
Numerical studies show that  for $N\geq 3$ there is a thermal 
 phase transition between confinement and deconfinement phases. 
Similar numerical studies detect a temperature $T_\chi$ at which
the broken chiral symmetry of $T=0$ QCD gives place to restored chiral symmetry
of high-$T$ QCD. The phase transition at $T_\chi$ is that of the chiral symmetry restoration (the lower plot in Fig.~(\ref{fig:surgery})).

In this case small-$L$ physics says little, if at all, about 
large-$L$ physics, our desired goal. We would like to create a different situation.
We would like to design a theory which (i) in the decompactification 
large-$L$ limit tends to conventional QCD and its QCD-like sisters;
(ii) at small $L$ is analytically tractable and has both confinement and chiral symmetry breaking;
and (iii) has as smooth transition between the small and large-$L$
domains as possible (the upper plot in Fig.~(\ref{fig:surgery})).  If this endeavor 
--- rendering small and large-$L$ physics continuously connected ---
is successful, we could try to
use small-$L$ laboratory to
extract lessons about QCD and QCD-like theories on $R_4$.

\begin{figure}[h]
 \centerline{\includegraphics[width=2in]{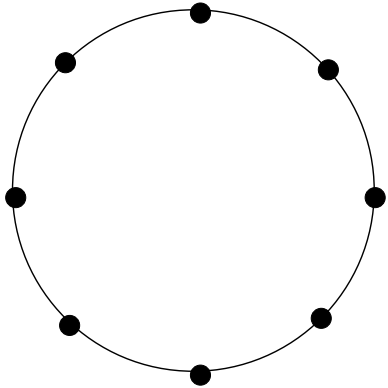}}
 \caption{\small $Z_N$ symmetric vacuum fields $v_k$.  }
 \label{zns}
 \end{figure}

 We will argue below that the goal can be achieved
by performing a so-called double-trace deformation of QCD and QCD-like 
theories.\footnote{The double trace deformations were previously  discussed  in the context of gauge/string theory dualities in \cite{Aharony:2001pa, Witten:2001ua, Berkooz:2002ug,Barbon:2002xk}, as well as
in field theory   \cite{Schaden:2004ah, Pisarski:2006hz, Myers:2007vc}.}  
To this end we add a non-local operator
 \begin{equation}
P[U({\bf x}) ]= \frac{2}{\pi^2 L^4} \,\sum_{n=1}^{\left[\frac{N}{2}\right]} d_n |
\, {\rm Tr}\, U^n({\bf x} )|^2  \qquad {\rm for} \;  {\rm SU}(N),
\label{fourma}
\end{equation}
to the QCD action,
\beq
\Delta S = 
\int_{R_3} d^3x\, L \,  P[U({\bf x}) ]\,,
\label{fivem}
\eeq
were $d_n$  are numerical parameters to be judiciously chosen.  
The theories obtained in this way will be labeled by  asterisk.
In minimizing $S+\Delta S$ the effect due to deformation
(\ref{fourma}) is two-fold. First, it tends to minimize $ |{\rm Tr}\, U({\bf x} )|$.
Second it tends to maximize the distance between the
eigenvalues of $U$.
It  is necessary to have a polynomial 
of order  $[N/2]$ to force the eigenvalues of the Polyakov line to be maximally apart from one another, i.e. to push the theory towards  the center-symmetric point
depicted in Fig.~\ref{zns}.
Here $[x]$ stands for the integer part of $x$. To stabilize the vacuum sufficiently close to the center-symmetric configuration the coefficients $d_n$ must be large enough,
presumably, $d_n\sim 1$. Some technical details are discussed in Appendix.

At large $L$, the deformation switches off and has 
no impact on the theory, i.e. QCD$^*\approx$ QCD. However, at small $L$
the impact is drastic. Given an appropriate choice
of $d_n$'s the  deformation (\ref{fivem}) forces the theory to pick up
the following set\,\footnote{
More exactly, the set of VEVs will be very close to (\ref{10}).}
of the vacuum expectation values (VEVs):
\beq
v_k = e^{\frac{2\pi i k}{N}},\qquad k=1,... , N,
\label{10}
\eeq
(or permutations), see Fig.~\ref{zns}.
If we define
\beq
e^{iaL} \equiv U, 
\eeq
\beq
a =\sum_{\rm Cartan\,\, gen} a_c T^c \equiv {\rm diag}\{ a_1, a_2, ..., a_N\}\,,\qquad  \sum_{k=1}^N a_k =0\,,
\label{4}
\eeq
it is obvious that Eq.~(\ref{10}) implies
\beqn
&& \{La_i\} = \{ -i L\, \ln v_i \,\, ({\rm mod}\,\, 2\pi)\}
\nonumber\\[3mm]
&&= \left\{-\frac{2\pi [N/2]}{N},\,\,  -\frac{2\pi ([N/2]-1)}{N}, ....,\,
\frac{2\pi [N/2]}{N}\right\}\,.
\label{12}
\eeqn
 This means, in turn, that
the theory is maximally Higgsed,
\beq
{\rm SU}(N)\to {\rm U}(1)^{N-1}
\label{higgsed}
\eeq
and weakly coupled at $L\ll \Lambda^{-1}$. The gauge bosons from the Cartan subalgebra
(to be referred to as photons) remain classically massless,
while the off-diagonal gauge bosons (to be referred to as $W$ bosons)
acquire large masses. The effective low-energy dynamics is that of compact QED. (See footnote~\ref{f6}, though.) 
It is not trivial. Dual photons acquire exponentially small  masses nonperturbatively
through the instanton-monopole mechanism \cite{P2,Unsal:2007vu}.
The mass gap generation in the dual description amounts to
linear Abelian confinement (at exponentially large distances). 
Chiral bifermion condensates are generated too \cite{Davies:2000nw,Unsal:2007jx}.
Thus, the dynamical patterns in the small and large-$L$ domains
do not seem to be that different from each other. Details are different (e.g. Abelian vs. non-Abelian confinement), but gross features appear to be similar.
It is not unreasonable to  expect that there is no phase transition in $L$.

What is meant when we speak of Abelian/non-Abelian confinement 
\cite{SY,Armoni:2007kd}?
In the former case the gauge group acting in the infrared (IR) and responsible for the flux tube formation is Abelian (i.e. U(1)$\times$U(1) ...). In the latter case
we deal with a non-Abelian group in the infrared.

 The best-known example exhibiting both regimes is the Seiberg--Witten solution
 \cite{Seiberg:1994rs} of a deformed
${\mathcal N}=2$ super-Yang--Mills theory. If the deformation parameter $\mu$ is small,
$$
|\mu |\ll \Lambda \,,
$$
the SU$(N)$ gauge group is spontaneously broken down to U(1)$^{N-1}$, and the  confining string  is a generalization of 
the  Abrikosov vortex \cite{ANO}.
In the opposite limit
$$
|\mu |\gg\Lambda \,,
$$
the breaking of SU$(N)$ down to U(1)$^{N-1}$ does not occur.
The infrared dynamics is determined by SU$(N)$; the corresponding flux tubes
should be non-Abelian. Since the theory is holomorphic  in $\mu$,
the Abelian and non-Abelian confinement regimes  are 
expected to be smoothly connected.

Another example which should be mentioned
(and which is close in formulation
to what will be presented below) where it is believed that  no phase 
transition in $L$ takes place    is $\N=1$  supersymmetric Yang--Mills
(SYM) theory on $R_3\times S_1$ \cite{Cohen:1983fd, Katz:1996th, Davies:1999uw, Davies:2000nw, Unsal:2007jx}. 

We expect that QCD$^*$ and QCD$^*$-like theories are of this type ---
there is no phase transition between the Abelian confinement small-$L$  and 
non-Abelian confinement large-$L$ domains. 

\vspace{2mm}

{\bf Conjecture:}  \label{claim}The deformed {\bf one}-flavor QCD-like
theories interpolate  from small $r(S_1)$ to large $r(S_1)$  without phase transitions.  

\vspace{2mm}

Since the theories under consideration are non-supersymmetric we cannot back up this statement by holomorphy.
Thus, the smoothness conjecture is on a somewhat weaker basis than 
in the  Seiberg--Witten problem. However, arguments to be presented below
can be viewed as at least some evidence in favor of the absence of the phase transition
in $L$. More evidence can (and should) be provided by lattice studies. 

In QCD-like theories with more than one flavor, chiral symmetry breaking
($\chi$SB) occurring on $R_4$ at strong coupling produces 
$N_f^2-1$ Goldstone mesons. Needless to say, it is impossible to get such Goldstones 
at weak coupling at small $L$.  However, if one 
considers theories with {\em one} fermion flavor in the center-symmetric
regime, there are no obvious reasons for a chiral phase transition.
The chiral symmetry in such theories is discrete, and its spontaneous breaking results
in domain walls rather than Goldstones. This phenomenon can show up both at strong and weak couplings. In this paper we will limit ourselves to QCD-like theories with a single flavor.

To be more exact, we will discuss in some detail SU$(N)$ Yang--Mills theory
with one fermion in the fundamental and two-index AS representations. Analysis of the 
 two-index S fermion essentially runs parallel to that of the AS case.
 We will also dwell on SU$(N)\times$SU$(N)$ Yang--Mills with
 the bifundamental fermion. The number of colors $N$ is
 {\em not} assumed to be large. The large-$N$ limit and the case of fermions in the
 adjoint representation were treated elsewhere
 \cite{Armoni:2007kd,Unsal:2007vu}.

Among other results, we will,
in particular, 
argue
 that many dynamical features  of SU$(N) \times$SU$(N)$ orbifold QCD 
are  remarkably close to those of SYM
theory. The pattern of the chiral symmetry breaking, the mass gap, the nonperturbative spectrum, the  $k$-string tensions ---  all of the above are demonstrated to 
coincide in these two theories. 

The paper is organized as follows. In Sect.~\ref{s2} we outline our formulation of the problem and briefly review general aspects of one-flavor QCD-like theories
on $R_4$ and $R_3\times S_1$. We also review dual description
of three-dimensional Yang--Mills (the Georgi--Glashow model),
and Polyakov's confinement. In Sect.~\ref{FF} we consider the case
of one fermion in the fundamental representation and solve the
theory at small $r(S_1)$. In Sect.~\ref{s4} we carry out the same analysis
in the SU$(N)\times$SU$(N)$ theory with one bifundamental fermion (orbifold theory).
In Sect.~\ref{s5} we consider Yang--Mills theory with
one fermion in the two-index antisymmetric representation of SU$(N)$.
Section~\ref{s6} is devoted to $\theta$ dependence.
In Sect.~\ref{plan} we discuss how our results are
related to planar equivalence.
Finally, Section~\ref{s7} summarizes our results and outlines some problems for future investigation.

\section{QCD and QCD-like theories on \boldmath{$R_4$} and \\
\boldmath{$R_3\times S_1$}: general aspects}
\label{s2}

 We will consider one-flavor QCD-like theories with the SU$(N)$ gauge group and 
 fermions in the following representations:
\begin{equation}
{\cal R} = \{\rm F, AS, S, Adj, BF \}  \,,
\label{Eq:classes}
\end{equation}
where F stands for fundamental, AS and S are two-index antisymmetric and symmetric representations,
Adj stands for adjoint, while BF for bifundamental. In all cases except Adj we deal with
the Dirac fermion field, while in the adjoint case with the Majorana (Weyl) spinor.
This is nothing but supersymmetric Yang--Mills (SYM) theory.
In the BF case the gauge group is  SU$(N)\times$SU$(N)$, with the fermion field
being fundamental with respect to the first SU$(N)$ and antifundamental with respect to the second SU$(N)$. 
For the adjoint fermions we will use the following nomenclature.
The theory with one Majorana flavor will be referred to as SYM,
while in the case of two or more flavors we will speak of QCD(Adj).

The boundary conditions for
fermions can be either periodic  $(\cal S^{+})$ or antiperiodic $(\cal S^{-})$ 
in the compactified dimension.
Yang--Mills theories with two-index fermions received 
much attention lately in connection with planar equivalence between such theories and
SYM theory (see \cite{ Armoni:2007vb} and references therein). At $N=3$ the AS theory  is equivalent to F.

 Theoretically the most informative is $\N=1$  SYM theory.
For periodic  spin connection   $\cal S^{+}$ this theory has unbroken center symmetry and broken discrete chiral symmetry  for any   $r(S_1)$. 
In fact, the chiral condensate $\langle {\rm Tr} \lambda\lambda\rangle$
was exactly calculated long ago
\cite{SVC,Davies:2000nw}, both on $R_4$ and $R_3\times S_1$,
and was shown to be totally independent of the value of $r(S_1)$.
More recently, this theory was demonstrated \cite{Unsal:2007jx} 
to possess Abelian confinement
at small $L$. Therefore, there is no {\em obvious}
obstruction for the $L$ evolution to be smooth. We know that at $L$ larger than 
the strong scale $\Lambda^{-1}$, the neutral sector observables in 
$\N=1$ SYM theory and QCD(AS/S/BF) are remarkably 
close and only differ by mild  $O(1/N)$ effects. However, the 
complex representation fermions break center symmetry at small $r(S^1)$ 
implying that these theories become drastically different from $\N=1$ SYM
theory. The double-trace deformation (\ref{fivem}) is designed
to maintain this similarity at small $r(S_1)$ too. One of the most intriguing 
findings of this paper is that the analytical tractability of $\N=1$ SYM 
theory in the small-$r(S_1)$ limit is not necessarily a consequence of supersymmetry.  
The unbroken center symmetry is equally important.   

Briefly summarizing our knowledge of other one-flavor QCD-like
theories\,\footnote{A part of this knowledge is folklore.} on $R_4$ 
we can say the following. All these theories are expected to exhibit:

\vspace{1mm}

(i)  Mass gap: there are no massless particles in the physical spectrum;

\vspace{1mm}

(ii) Non-Abelian confinement: 
the gauge group is not Higgsed, chromoelectric flux tubes are formed between
quarks and antiquarks,
these flux tubes are not stable, generally speaking, since the dynamical quark
pair production can break them. No color-charged objects are present in the physical spectrum;

\vspace{1mm}

(iii)  Discrete chiral symmetry breaking\,\footnote{ For F representation, the 
anomaly-free $Z_2$ is  the fermion number and can not be spontaneously 
broken.  The theory has    a unique vacuum.}  for  $ {\cal R} = \{\rm AS, S, BF, SYM \}$:
The one-flavor QCD-like theories on $R_4$ possess an axial U(1) symmetry at the classical level. Only a discrete subgroup of it,  $Z_{2h}$,  is the symmetry  of the quantum theory, 
\begin{equation}
Z_{2h}= \{ Z_2, Z_{2N-4},   Z_{2N+4},  Z_{2N},  Z_{2N} \}    \;\;  {\rm for} \;\;
  {\cal R} = \{\rm F, AS, S,BF,SYM \} ,
\end{equation}
respectively. Here  $2h$ is the number of the fermion 
zero modes in the instanton background.
In all cases but F the axial $Z_{2h}$ is spontaneously broken down to $Z_2$.
 Discrete symmetry breaking, unlike that of the continuous symmetries, does not lead to Goldstone bosons.  
Instead, the 
theory must possess $h$ isolated vacua. 

The above picture follows from multiple lattice calculations, and 
supersym\-metry-based and large-$N$ methods.

In this work the  double-trace deformation of QCD(${\cal R}$) on 
$S_1 \times R_3$ with small $r(S_1)$ is used to stabilize the 
theories under consideration at (or, more exactly, very close to)
a center-symmetric point. 
At small $r(S_1)$  the non-Abelian gauge group
is Higgsed down to the maximal Abelian subgroup,
but neither confinement nor the above chiral properties
are lost. We will explicitly demonstrate confinement,
the discrete chiral symmetry breaking, and  mass gap generation.
 
On  $S_1 \times R_3$ the Yang--Mills Lagrangian
with one fermion flavor in 
the representation ${\cal R}$
takes the form
\begin{equation}
S= \int_{R_3 \times S_1} \frac{1}{g^2} \left[ \frac{1}{2}\,  \tr  \,
F_{MN}^2 (x)  + i
 \bar  \Psi  \Dslash 
 \Psi  \right]
\label{eq:cont}
\end{equation}
where $\Psi$  is the four-dimensional Dirac spinor in the
representation ${\cal R}= \{{\rm F, AS,S } \}$  of the  gauge group  SU$(N)$,  $F_{MN}$ is the non-Abelian gauge field strength,\footnote{Throughout the paper we use 
the following notation:  $M,\, N=1,\, \ldots,\,  4$ are
four-dimensional Lorentz indices while  and  ${\mu, \nu}=1, 2, 3$ are 
three-dimensional indices. We normalize the Lie algebra generators as ${\rm Tr} \;  t^A t^B = \frac{1}{2} \delta^{AB}$. } and $\Dslash=\gamma_M D_M= \gamma_M( \partial_M + i A_M)$ is the covariant derivative acting on representation  ${\cal R}$. For  QCD(BF), the gauge group is SU$(N) \times {\rm SU}(N)$ and 
gauge field part of the action must be replaced by 
$$F_{MN}^2 (x) \rightarrow F_{1, MN}^2 (x) + 
 F_{2, MN}^2 (x)\,.$$
In this theory the fermion is in the bifundamental  representation. 
In terms of its Weyl components, the Dirac   fermions are decomposed as 
\begin{equation}
\Psi= \left(\begin{array}{l} 
                   \lambda \cr 
                   \bar \psi
                    \end{array}
 \right)
\end{equation} 
where $\lambda,\,\,  \psi$ are two-component (complex) Weyl
spinors. In three dimensions $\lambda,\,\,  \psi$ represent two
Dirac spinors.

We must use the Kaluza-Klein (KK) mode decomposition
for all fields in the Lagrangian. If we discard all modes other than zero we will arrive at
a three-dimensional theory with a gauge field, a scalar field in the adjoint
and two three-dimensional spinors. 
The $S_1 \times R_3$ reduction of $R_4$ Yang--Mills does not quite  lead to 
three-dimensional Yang--Mills, but at first, we will ignore this nuance, to be discussed
in detail later, and will briefly review the phenomena that
occur in three-dimensional Yang--Mills with a scalar field in the adjoint (discarding fermions for the time being).

Long ago Polyakov considered three-dimensional SU(2) Georgi--Glashow mo\-del 
(a Yang-Mills + adjoint Higgs system)  
in the Higgs regime \cite{P2}. In this regime SU(2) is broken down to U(1),
so that at low energies the theory reduces to compact electrodynamics. 
The dual photon is a scalar field $\sigma$ of the phase type
(i.e. it is defined on the interval $[0, 2\pi ]$):
\beq
F_{\mu\nu}
 =\frac{g_3^2}{4\pi} \, \varepsilon_{\mu\nu\rho}\left( \partial^\rho\,\sigma\right)\,,
 \label{14.8}
\eeq
where  $g_3^2$  is the three-dimensional  gauge coupling with mass dimension
  $ [g_3^2]=+1$. 
 In perturbation theory the dual photon $\sigma$
is massless.   However, it acquires a mass
due to instantons (technically, the latter are identical to the 't Hooft--Polyakov
monopoles, after the substitution of one spatial dimension by imaginary time;
that's why below we will refer to them as to the instantons-monopoles).
In the vacuum of the theory,  one deals with a gas  of instantons interacting
according to the Coulomb  law.   
The dual photon mass is due to the Debye screening.
In fact, the dual photon mass is determined by the one-instanton vertex,
\beq
m_\sigma \sim m_W^{5/2} g_3^{-3}e^{-S_0/2}
\eeq
where $S_0$ is the one-instanton action,
\beq
S_{0} = 4\pi \, \frac{m_W}{g^2_3}\,,
\label{14.5}
\eeq
$m_W$ is the lightest $W$ boson mass, see below. In terms of four-dimensional quantities
$S_0 = 8\pi^2/(Ng^2)$. 
As a result, the low-energy theory
is described by a three-dimensional sine-Gordon model,
\beq
{\cal L}_\sigma = \frac{g_3^2}{32\pi^2} (\partial_\mu\sigma )^2 + c_1 
m_W^5g_3^{-4}e^{-S_0}
 \, \cos\sigma\,. 
 \label{ptpm}
\eeq
where $c_1$ is an undetermined prefactor. 
The coefficient in front of $ e^{-S_0}
 \, \cos\sigma$,
$$
\mu \equiv c_1 
m_W^5g_3^{-4}  \,,
$$ 
has  mass dimension is $[\mu ]= +3$. The combination $\mu e^{-S_0}$ is the monopole 
fugacity. 

This model supports a domain line\,\footnote{Similar to the axion domain wall.} (with 
$\sigma$ field vortices at the endpoints)
which in 1+2 dimensions must be interpreted as a string.
Since the $\sigma$ field  dualizes three-dimensional photon, the $\sigma$ field vortices
in fact represent electric probe charges in the original formulation, connected by the
electric flux tubes which look like domain lines in the dual formulation. 

Now, if we switch on massless adjoint fermions, as  in \cite{Affleck:1982as},
the mass gap generation does not occur in the Polyakov model {\em per se}. 
This is due to the fact that the instanton-monopoles acquire fermion zero modes
which preclude the potential term as in Eq.~(\ref{ptpm}). 
Correspondingly, the dual photons remain massless
and the model no longer supports domain lines.
The linear confinement is gone.

This situation, changes, however, if three-dimensional Yang--Mills theory is
obtained as a low-energy reduction of a four-dimensional gauge theory 
on $S_1 \times R_3$ with small $r(S_1)$. 
When the adjoint Higgs field is compact, as in Fig.~\ref{zns}, in addition to $N-1$ 
't Hooft--Polyakov monopole-instantons there is one extra monopole   
 (whose existence is tied up to $\pi_1 (S_1) \neq 0$). It can  be referred to as
the Kaluza--Klein (KK) monopole-instanton.\footnote{The eigenvalues 
shown in Fig.~\ref{zns} may be viewed as Euclidean D2-branes. $N$ split branes support a spontaneously broken U(1)$^{N}$ gauge theory, whose  U(1) 
center of mass decouples, and the resulting theory is U(1)$^{N-1}$. The $N-1$ 
't Hooft--Polyakov monopoles may be viewed as Euclidean D0 branes 
connecting the eigenvalues $(a_1 \rightarrow  a_2), \,
(a_2 \rightarrow  a_3),\,  \ldots, \, (a_{N-1} \rightarrow a_{N})$. Clearly, 
we can also have a monopole which connects $(a_N \rightarrow  a_1)$ which owes its existence to the periodicity of the adjoint Higgs field,  or equivalently,  to the fact that the underlying theory is on $S_1 \times R_3$.  Usually it is  called the KK monopole.  The 
Euclidean D0 branes with the
opposite orientation, connecting $(a_{j} \leftarrow a_{j+1}),\,\, j=1, \ldots 
N $, are the antimonopoles. This viewpoint makes manifest 
the fact that the KK  and 't Hooft--Polyakov monopoles are all on the same footing.  The magnetic and topological charges of the monopoles 
connecting  $(a_{j} \leftrightarrow a_{j+1}) $ is 
$\pm \Big( (4\pi/g)\mbox{ \boldmath $\alpha$}_j, \frac{1}{N} \Big)$ 
where the direction of the arrow is correlated with the sign of the charges. 
}
Each of these monopoles carries  fermion zero modes, hence they cannot contribute to the bosonic  potential at the level $e^{-S_0}$. They can and do
contribute at the level $e^{-2S_0}$. 

Indeed,
the bound state of the 't Hooft--Polyakov monopole-instanton with magnetic charge  
 $\mbox{\boldmath $\alpha$}_i$ and anti-monopole with charge  $-\mbox{\boldmath $\alpha$}_{i+1}$  has no fermion zero modes: its topological charge
coincides with that of the perturbative vacuum. Hence, such a bound state 
can contribute to the bosonic potential. Let  
\beq
\Delta^{0}_{\rm aff}= \{ 
 \mbox{\boldmath $\alpha$}_1,  \mbox{\boldmath $\alpha$}_2, \ldots,  \mbox{\boldmath $\alpha$}_N \}
 \end{equation}
 denote the extended (affine)  root system of SU(N) Lie algebra.   
If we normalize the magnetic and topological  charges of the monopoles  as 
 \begin{equation}
\left(  \int_{S^2} F,   \int 
 \frac{g^2}{32 \pi^2}  \, F_{\mu \nu}^a {\widetilde F}^{\mu \nu\,,a} \right)  = \left(  \pm  \frac{4 \pi}{g} 
\mbox{\boldmath $\alpha$}_i,  \pm   \frac{1}{N}  \right),   \quad {\rm for} \; \; \mbox{\boldmath $\alpha$}_i \in \pm \Delta^{0}_{\rm aff} 
\label{38pp}
 \end{equation} 
where $ \mbox{\boldmath $\alpha$}_i$ stands for the simple roots of the affine Lie algebra
then the following bound states are relevant:
\beq
\left[ \frac{4 \pi}{g} \mbox{\boldmath $\alpha$}_i,\,\, \frac{1}{N} \right]  + \left[ -\frac{4 \pi}{g}  \mbox{\boldmath $\alpha$}_{i+1}, \,\,- \frac{1}{N} \right] =\left[\frac{4 \pi}{g} \left( \mbox{\boldmath $\alpha$}_i - \mbox{\boldmath $\alpha$}_{i+1}
\right), \,\, 0 \right].
\label{38}
\eeq
This pair is  stable,  as was shown in Ref.~\cite{Unsal:2007vu}, where it is referred 
to as a magnetic bion. Thus,  we can borrow Polyakov's discussion of magnetic monopoles and apply directly to these objects.   The magnetic bions  will induce a mass term  for the dual photons via the Debye screening, the essence of Polyakov's mechanism. 
 
The vacuum field (\ref{12}) of the deformed SU$(N)$ theory respects  the  (approximate) center  symmetry $Z_N$. This field configuration  breaks the gauge symmetry as indicated in 
(\ref{higgsed}).
Due to the gauge symmetry breaking, electrically charged particles acquire masses. 
(By electric charges we mean charges with regards to $N-1$ 
``photons" of the low-energy theory.)
The set of $N-1$ electric charges and masses of $N$ lightest $W$ bosons are 
\begin{eqnarray}
\mbox{\boldmath $q$}_{W_{\mbox{\boldmath $\alpha$}}} = g \mbox{\boldmath $\alpha$}\,, \qquad  m_{W_{\mbox{\boldmath $\alpha$}}}=  
\frac{2 \pi  }{N L} \,,
\end{eqnarray}
where $\mbox{\boldmath $\alpha$}_i$ ($ i=1, ... , N$) are the simple and affine roots of the SU$(N)$ Lie algebra (see Eq.~(\ref{dop6})). Note that 
$N$ lightest $W$ bosons are degenerate in the center-symmetric vacuum. The remaining    
$N^2 -N$ charged $W$ bosons can be viewed as composites of the above.

The stabilizing double-trace term (\ref{fourma}) contributes to the self-interaction
of the physical (neutral) Higgs fields. Assuming that all coefficients $d$
are of order one, the masses of these fields are ${\mathcal O}(g/L)$.
For instance, for SU(2) and SU(3) the physical Higgs masses are $(g\sqrt{d_1})/L$.
These masses are much lighter than those of the $W$ bosons but much heavier
than those of the fields in the effective low-energy 
Lagrangian (dual photons, see Eq.~(\ref{40}) below).
The stabilizing double-trace term (\ref{fourma}) also contributes
to corrections to the $W$ boson masses. They are expandable in $g^2$,
i.e.
$$
m_{W_{\mbox{\boldmath $\alpha$}}}=  
\frac{2 \pi  }{N L} \left(1 +{\mathcal O}(g^2)
\right).
$$

In the SU$(N)$  gauge theory with an  adjoint fermion on $R_3\times S_1$, which
is Higgsed according to (\ref{higgsed}), the bosonic part of the effective low-energy Lagrangian is generated by the pairs (\ref{38}), and hence the potential is proportional to $e^{-2S_0}$, rather than $e^{-S_0}$ of the Polyakov problem. 
If we introduce an $(N-1)$-component  vector $\mbox{\boldmath $\sigma$}$,
\beq 
\mbox{\boldmath $\sigma$} \equiv \left(\sigma_1, ...., \sigma_{N-1}\right), 
\eeq 
representing $N-1$ dual photons
of the $U(1)^{N-1}$ theory, the bosonic part of the effective Lagrangian can be written as
\beq
{\cal L}(\sigma_1, ...., \sigma_{N-1}) =  \frac{g_3^2}{32\pi^2} (\partial_\mu\mbox{\boldmath $\sigma$} )^2 + 
 c m_W^6 g_3^{-6}e^{-2S_0} 
 \, \sum_{i=1}^{N} \cos \left( \mbox{\boldmath $\alpha$}_i - \mbox{\boldmath $\alpha$}_{i+1}\right)\mbox{\boldmath $\sigma$}\,,
 \label{40}
\eeq
where $c$ is an undetermined coefficient
and $g_3$ is the three-dimensional coupling constant,
\beq
g_3^2 = g^2\, L^{-1}\,.
\eeq
In terms of four dimensional variables, the  magnetic bion fugacity is 
\beq
 m_W^6 g_3^{-6}e^{-2S_0} \sim m_W^3 g^{-6} e^{-2S_0}
\eeq
We remind  that $\mbox{\boldmath $\alpha$}_i$ ($ i=1, ... , N-1$) represent the magnetic charges of $(N-1)$ types of the
't Hooft--Polyakov monopoles while the affine root 
\beq
\mbox{\boldmath $\alpha$}_N= -\sum_{i=1}^{N-1} \mbox{\boldmath $\alpha$}_i
\label{dop6}
\eeq
 is the magnetic charge 
of the KK monopole.  
 Note that the bion configurations  that contribute to the effective Lagrangian 
have magnetic charges $\mbox{\boldmath $\alpha$}_i - \mbox{\boldmath $\alpha$}_{i+1}$ and vertices  $e^{i(\mbox{\boldmath $\alpha$}_i - \mbox{\boldmath $\alpha$}_{i+1}) \mbox{\boldmath $\sigma$}}$,  corresponding to a  product of a  monopole vertex  
 $e^{i\mbox{\boldmath $\alpha$}_i \mbox{\boldmath $\sigma$}}$
with charge  $\mbox{\boldmath $\alpha$}_i$,  and   antimonopole 
vertex $e^{-i\mbox{\boldmath $\alpha$}_{i+1} \mbox{\boldmath $\sigma$}}$  
with charge $-\mbox{\boldmath $\alpha$}_{i+1}$ (without the zero mode insertions).  With the $Z_N$-symmetric vacuum 
field  (\ref{12}) all fugacities  are equal.   

Equation (\ref{40}) implies that nonvanishing masses
proportional to $e^{-S_0}$ are generated for all $\sigma$'s. They are much smaller 
than the masses in 
the Polyakov model in which they are $\sim e^{-S_0/2}$. 

There are
$N-1$ types of Abelian strings (domain lines). Their tensions are equal to each
other and proportional to $e^{-S_0}$. Linear confinement develops
at distances larger than $e^{S_0}$.

Needless to say,  the physical spectrum
in the Higgs/Abelian confinement regime is richer than that
in the non-Abelian confinement regime. If in the latter case
only color singlets act as asymptotic states, in the Abelian confinement regime
all systems that have vanishing $N-1$ electric charges have finite mass and represent
asymptotic states.

{\bf Note 1:} For SU(2) and SU(3) Yang--Mills theories, the double-trace deformation is a particularly simple  monomial 
\begin{equation}
P[U({\bf x}) ]=  \frac{2}{\pi^2 L^4} \ d_1 |\tr\,  U({\bf x} )|^2  \quad {\rm for} \;  {\rm SU}(2),\,\,\,  {\rm SU}(3) \,.
\end{equation}

{\bf Note 2:}
One can  be concerned that the deformation potential is
given  in terms of multi-winding line operators, 
and looks nonlocal. 
In the $L \Lambda\ll1 $  region where the  deformation is crucial, there  is no harm in viewing the 
deforming operator as ``almost local" since we are concerned with physics at scales much larger than the compactification scale. 
In the decompactification limit  where the  deformation is indeed nonlocal, it is not needed since its dynamical role is negligible. If one wants to be absolutely certain, one
can insert a
filter function
as the coefficient of the double-trace operator which shuts it off exponentially   
$\sim e^{- L^2 \Lambda^2}$  at large $L$ in order not to deal with a 
non-local theory.  

\section{ QCD with one  fundamental fermion}
\label{FF}

QCD(F) on $R_4$ possesses a U(1)$_V \times {\rm U}(1)_A $ symmetry, at the classical level acting as 
$$
\Psi \rightarrow e^{i \alpha } \Psi,\quad \Psi \rightarrow e^{i \beta \gamma_5 } \Psi\,.
$$ 
Due to nonperturbative effects, only the anomaly-free $Z_2$ subgroup of the U(1)$_A$ is the genuine axial symmetry of the theory, the fermion number mod
two. This symmetry is already a part of the vector U(1$)_V$ symmetry,  and, hence,
cannot be spontaneously broken. However, a bifermion condensate (which does not break any chiral symmetry) is believed to exist on $R_4$ as well as on  $S_1 \times R_3 $  with sufficiently large $r(S_1)$.   

The microscopic  QCD Lagrangian also possesses the discrete symmetries  $C, P, T $, and 
continuous three-dimensional  Euclidean  Lorentz symmetry  SO(3).  Thus, the  symmetries of the original theory are 
\begin{eqnarray}
{\rm U}(1)_V  \times  C \times P \times T \,.
\label{Eq:allsymF}
\end{eqnarray}
The double-trace deformation respects all these symmetries. (Otherwise this would 
explicitly contradict the claim made in Sect.~\ref{claim}.)  
Below,  we will construct a low-energy effective theory  QCD(F)* assuming
that the double-trace terms stabilize the theory 
in the center-symmetric vacuum.  As usual, the set of all possible 
operators that can appear in the effective 
low-energy theory is restricted by the underlying symmetries (\ref{Eq:allsymF}). 

Integrating out weakly coupled KK modes with nonvanishing frequencies 
$$
|\omega_n| \geq  \frac{2 \pi n}{L}\,, \quad n \neq 0\,,
$$ 
and adding the stabilizing 
deformation term (\ref{fourma})
to the QCD(F) Lagrangian, we obtain the QCD(F)* theory.  
This is the Yang--Mills + {\it compact} adjoint  Higgs system 
with  fundamental  fermions on $R_3$. 

The  action is\,\footnote{
Our four-dimensional  Dirac $\gamma $ matrix conventions 
are 
$$
\gamma_{M}= \{ \gamma_{\mu} ,\,\,  
\gamma_{4} \}\,,\quad \gamma_{\mu} = \sigma_1 \otimes \sigma_{\mu}\,, \quad \gamma_{4} = \sigma_2 \otimes I\,.$$ 
With this choice, the Dirac algebras in four and three dimensions are 
$\{\gamma_M, \gamma_N \} = 2 \delta_{MN}$ and  $\{ \sigma_{\mu} , \sigma_{\nu} \}= 2 \delta_{\mu \nu}$.   It will be convenient to define $\bar \sigma_M= (\sigma_{\mu}, -i I) \equiv  (\sigma_{\mu}, \sigma_4) $ and  
$\sigma_M= (\sigma_{\mu}, i I) \equiv  (\sigma_{\mu}, - \sigma_4) $\,. }

\begin{eqnarray}
S=&& \int_{R_3}  \;  \frac{L}{g^2}  \Big[ {\rm Tr} \Big(  \frac{1}{2} F_{\mu \nu}^2 + 
 (D_{\mu} \Phi)^2   + g^2  V [\Phi] \Big)  \nonumber\\[3mm]  
 && +  i \bar \lambda 
\left( \sigma_{\mu} (\partial_{\mu} + i A_{\mu}) + i \sigma_4 \Phi \right)  \lambda    
\nonumber\\[3mm]  
 && + 
  i \bar \psi
\left( \sigma_{\mu} (\partial_{\mu} - i A_{\mu}) - i \sigma_4 \Phi \right)  \psi  
    \Big]  \,, 
\end{eqnarray}
where  $\psi$ and $\lambda$ are the two-component three-dimensional  
Dirac spinors  which arise upon 
reduction of the  four-dimensional Dirac spinor $\Psi$.  Note that  $\lambda$ and $\psi$ has opposite gauge charges, where $\lambda$ and $\bar \psi $ are fundamental and  
$\bar \lambda$ and $\psi $ are anti-fundamental. As usual, in Euclidean space, there is no relation between barred and unbarred variables, and they are not related to each other by conjugations.

The potential $V [\Phi] $ which is the sum of the one-loop potential and deformation potential  has its minimum  located at (\ref{10}) (or (\ref{12})).
The fermion contribution to the effective one-loop potential involves terms such as 
$\tr \, U + \tr\,  U^{*}$.  These terms explicitly break the $Z_N$ center symmetry  and  slightly shift the position of the eigenvalues of $\langle U \rangle $
from the minimum (\ref{10}). 
However, this is a negligible  $\O(g/d_n)$ effect suppressed by a judicious choice of the deformation parameters. Hence, we neglect this effect below.\footnote{If the  eigenvalues are separated not equidistantly, yet the separations are nonvanishing for any pair, the gauge symmetry breaking SU$(N) \rightarrow$ U(1)$^{N-1}$ still takes place. In the nonperturbative analysis below, this fact manifests itself as an unequal action (or fugacity) for different types  of monopoles.  The analysis in this latter case will not be qualitatively different.}
 
There are $N-1$ distinct U(1)'s in this model, corresponding to $N-1$ distinct electric charges. If we introduce a quark $\Psi $  in the fundamental representation of 
SU$(N)$ each component $\Psi_i$ ($i=1, ... , N$) will be characterized by a set
of $N-1$ charges, which we will denote by $\mbox{\boldmath $q$}_{\Psi_i}$,
\beq
\mbox{\boldmath $q$}_{\Psi_i} = g\, \mbox{\boldmath $H$}_{ii}\,,\quad
i=1, ... , N\,,
\eeq
where $\mbox{\boldmath $H$}$ is the set of $N-1$ Cartan generators.

All fundamental fermions, but two (one of each type $\psi$ and $\lambda$), acquire masses due to gauge symmetry breaking. These masses
are of order of $2\pi/L$ and depend on 
whether periodic or antiperiodic boundary conditions are imposed.
The fermions that remain massless in perturbation theory are the ones 
corresponding to the vanishing (mod $2 \pi$)  eigenvalue of the algebra-valued compact 
Higgs field $\Phi$, see Eq.~(\ref{12}) (equivalently, $v=1$, see Eq.~(\ref{10})).  

Thus, the low-energy effective Lagrangian includes $N-1$ photons and two fermions. 
Their interactions (in particular, an induced mass gap)
must arise due to nonperturbative effects.\footnote{\label{f6} It is important to distinguish this theory from the case of the noncompact adjoint  Higgs field, which is 
the Polyakov model with massless (complex representation) fermions. 
Both theories have identical gauge symmetry breaking patterns: 
SU$(N) \rightarrow {\rm U}(1)^{N-1}$. In perturbation theory, 
both theories reduce  (by necessity) to compact QED$_3$ with fermions.   
However, it is possible to prove that the  latter theory lacks confinement
since photons remain massless nonperturbatively. 
 This implies that if the symmetries at the cut-off scale 
are not specified, the question of confinement in compact QED$_3$ 
with massless fermions is ambiguous. The issue will be further discussed in a separate publication.} 

\subsection{Nonperturbative effects and the  low-energy \\
Lagrangian}

Nonperturbatively, there exist  topologically stable, semiclassical 
field configurations --- instantons-monopoles.
If the adjoint Higgs field were noncompact,   there would be $(N-1)$ types 
of fundamental monopoles.  There is, however,  an extra  KK monopole which 
arises due to the fact that the underlying  theory is  formulated  on a cylinder, 
$R_3 \times S_1$, or simply speaking,  $\Phi ({\bf x})$ is compact.  
The magnetic and topological  charges  of the  (anti)monopoles associated with root 
$\mbox{\boldmath $\alpha$}_i$  are given Eq.~(\ref{38pp}).

As follows from the explicit zero mode constructions of Jackiw  and Rebbi
\cite{Jackiw:1975fn} and the Callias index theorem \cite{Callias:1977kg}, there are 
two fermion zero modes localized on one of the $N$ constituent monopoles. 
Van Baal et al. demonstrated  \cite{ Bruckmann:2003ag, Chernodub:1999wg,GarciaPerez:1999ux, Bruckmann:2007ru} that as the boundary conditions of fermions  vary 
in the background with nontrivial holonomy,  the zero modes hop from a monopole to the  next one. With fixed boundary conditions, they are  
localized, generally speaking,  on a particular monopole.\footnote{
More precisely, the Callias index applies to $R_3$.  We need an   
 index theorem for the Dirac operators in the background of monopoles on $R_3 \times S_1$. Such a generalization of the Callias index theorem was carried out  in the work  of Nye and Singer \cite{Nye:2000eg}. For a clear-cut lattice realization of the fermion zero modes  explicitly showing on which monopole they are localized, see 
Ref.~\cite{Bruckmann:2003ag}.  }

The above implies that 
one of the monopole-induced vertices has two fermion insertions (the one on which
the fermion zero modes are localized)  and other $N-1$ elementary monopoles have no
fermion insertions (at the level $e^{-S_0}$). 
The set of the  instanton-monopole induced vertices can be summarized as 
follows:
\begin{equation}
\left\{  e^{-S_0} e^{i \, \mbox{\boldmath $\alpha$}_1 \mbox{\boldmath $\sigma$}} \,\lambda \psi, \qquad  e^{-S_0}e^{i \, \mbox{\boldmath $\alpha$}_j \mbox{\boldmath $\sigma$}} \, , \;\; 
 j=2, \ldots N \right\}\,,
\end{equation}
plus complex conjugate for antimonopoles. Thus, the leading 
nonperturbatively induced interaction terms in the effective Lagrangian are
\begin{eqnarray}
&& S^{\rm {QCD(F)}^*} =     \int_{R_3} \;   \Big[\,  
\,
 \frac{g_3^2}{32\pi^2} (\partial_\mu\mbox{\boldmath $\sigma$} )^2  +   \frac{1}{g_3^2}
i \bar \Psi \gamma^{\mu}( \partial_{\mu} +  i \mbox{\boldmath $q$}_\Psi \mbox{\boldmath $A$}_{\mu} )  \Psi 
\nonumber\\[3mm]
&& +  e^{-S_0} \, \Big( \tilde{\mu}\,   e^{i \, \mbox{\boldmath $\alpha$}_1 \mbox{\boldmath $\sigma$}} \,\lambda \psi  + 
     \mu\,  \sum_{\mbox{\boldmath $\alpha$}_j\in (\Delta^{0}_{\rm aff} - \mbox{\boldmath $\alpha$}_1 )}
             e^{i \mbox{\boldmath $\alpha$}_j\mbox{\boldmath $\sigma$} } 
             + {\rm H.c.}
  \Big)       \Big]  \,,
      \label{Eq:dQCD(F)}
\end{eqnarray}
where $\tilde{\mu}$ is  dimensionless constant. Note the non-canonical normalization of the 
bosonic and fermionic terms. This choice for fermions will ease the derivations of certain four physical quantities.   It is clearly seen that 
in the infrared description of QCD(F)*, we must deal not only with the dual photons, 
but also with electrically charged fermions.  

The three-dimensional effective  Lagrangian respects the symmetries (\ref{Eq:allsymF})  
of the microscopic (four-dimensional) theory.  In particular, the fermion bilinears such as 
$\bar \lambda \lambda$ (allowed by U(1)$_V$ and the Lorentz symmetry of the three-dimensional theory) are noninvariant under parity (see Appendix in 
Ref.~\cite{Affleck:1982as})
and, hence,  cannot be generated. 
On the other hand, $\langle \lambda \psi\rangle \neq 0$
can and is generated.
One can check that up to order $e^{-2S_0}$,  the Lagrangian 
(\ref{Eq:dQCD(F)}) includes all possible operators allowed by the symmetries (\ref{Eq:allsymF}). 

In the above Lagrangian, all operators are relevant in the renormalization-group sense.
The fugacity has mass dimension $+3$. If the kinetic term for fermion is canonically 
normalized, the  covariant photon-fermion interaction and  
instanton-monopole-induced term with the fermion  insertion has dimension  $+1$. 
Which operators will dominate the IR physics?
The answer to this question requires a full renormalization-group analysis of all 
couplings.  A preliminary investigation (along the lines of Ref.\cite{hermele-2004-70})  
shows that quantum corrections in the running of the couplings are tame and 
do not alter the fact that the instanton-monopole vertex terms are the most relevant 
in the IR of QCD(F)*. 

The $N-1$ linearly independent instanton-monopole vertices render 
all the  $N-1$ dual photons massive, with masses proportional to $e^{-S_0/2}$. Thus, the dual scalars are pinned at the bottom of the potential 
\beq
\mu\, 
e^{-S_0} \sum_{j=2}^{N} \cos \mbox{\boldmath $\alpha$}_j \mbox{\boldmath $\sigma$}\,.
\eeq
As a result, the would-be massless fermions will also acquire a mass term
of the type
\begin{equation}
\tilde\mu\, e^{-S_0} \; \lambda \psi \, .
\end{equation}
 The  fermion  mass is proportional to $e^{-S_0}$. Hence it is exponentially smaller than  
 the dual photon mass $\sim e^{-S_0/2}$.
Note that the fermion mass term is not associated with the spontaneous breaking of chiral symmetry.  This 
circumstance, as well as the hierarchy of mass between the photon and fermion,  is specific to one {\em fundamental} fermion and will change in the case of 
the two-index fermions. 

Since all $N-1$ dual photons become massive,
a probe quark $Q_i$ of every type $(i=1,...,N$) will be connected to its antiquark
by a domain line/string with the tension \footnote{This is also similar to the axion domain wall.}
\beq
T\sim g_3\, \mu^{1/2}\,  e^{-S_0/2}\,.
\eeq
The string between $Q_1$ and $\overline{Q_1}$ is easily breakable due to
pair production of $\lambda$'s and $\psi$'s.
In other words, the external charge $Q_1$ will be screened by the 
dynamical fermions with charge  $ \mbox{\boldmath $q$}_{\Psi_1}$. 
The strings between $Q_i$ and $\overline{Q_i}$ (with $i=2, \,...\, N$)
can break with an exponentially small probability due to pair creation
of the KK modes of $\Psi_i$. 
This amounts, of course, to the conventional statement about
large Wilson loops $C$, 
 \beqn
 &&
\Big \langle  \frac{1}{N} \Tr W(C) \Big \rangle \sim \frac{1}{N}
\sum_{i=1}^{N}  \Big \langle    e^{i  \int_C \mbox{\boldmath $H$}_{ii}  \mbox{\boldmath $A$} }
\Big \rangle 
\nonumber\\[3mm]
&&
= \frac{1}{N} e^{- \kappa P(C) } + \left(1-\frac{1}{N}\right)   e^{-T {\rm Area} (\Sigma)} 
\,.
 \eeqn
where $\kappa$ is the coefficient of the perimeter law,  $P(C)$ is the perimeter of the loop $C$, the  boundary of a surface $\Sigma$.
  
 \vspace{2mm}
 
{\bf Remark:} The product of the instanton-monopole-induced vertices is proportional to  the  Belyavin--Polyakov--Schwarz--Tyupkin (BPST) 
four-dimensi\-onal instanton vertex
\cite{BPST}, 
\begin{eqnarray}
&&
\left(e^{-S_0} e^{i \,\mbox{\boldmath $\alpha$}_1 \mbox{\boldmath $\sigma$}} \lambda \psi \right) \,\, 
\prod_{j=2}^{N}   \left( e^{-S_0} e^{i \,\mbox{\boldmath $\alpha$}_j \mbox{\boldmath $\sigma$}}
\right) 
\nonumber\\[4mm]
&&\sim   
e^{- \frac{8 \pi^2}{g^2}}\,\,  \bar \Psi( 1+ \gamma_{5} ) \Psi 
\,\, \exp \left( {i\sum_{i=1}^{N} \,\mbox{\boldmath $\alpha$}_i\, \mbox{\boldmath $\sigma$}} \right)= 
e^{- \frac{8 \pi^2}{g^2}} \,\,\bar \Psi( 1+ \gamma_{5} ) \Psi \,.
\label{36}
\end{eqnarray}  
This is consistent with the fact that the instanton-monopoles can be viewed as 
the BPST instanton constituents.
In Eq.~(\ref{36}) we used  the fact that the sum of the $N$ constituent 
instanton-monopole actions is in fact
the BPST instanton action, and the sum of the magnetic and topological charges 
of the constituent monopoles gives the correct quantum numbers of the BPST $R_4$  instanton, 
 \begin{eqnarray}
\sum_{i=1}^{N} \left( \int F \,,  \,\,\frac{g^2}{32 \pi^2} \,  F_{\mu \nu}^a {\widetilde F}^{\mu \nu\,,a}\right)_{i}  \,\,= (0, 1) \,,
\label{fracins}
\end{eqnarray}  
see Eq.~(\ref{38pp}).

\subsection{Bifermion condensate}

As stated earlier, one-flavor QCD formulated on $R_4$ has no chiral symmetry whatsoever.   The axial anomaly reduces the classical U(1)$_A$ symmetry to $Z_2$. A bifermion condensate exists and breaks no chiral symmetry.  
We can evaluate the value of the chiral condensate in QCD(F)* in the small $r(S_1)$ regime.  At large $r(S_1)$ (strong coupling) we know, from volume independence, that the condensate must get a value  independent of the radius.  
Let  $b_0$ denote  the leading coefficient of the $\beta$ function divided by $N$,  
\begin{equation}
b_0= \frac{1}{N} \left.\left(\frac{11N}{3} - \frac{2N_f}{3}\right)\right|_{N_f=1} = \frac{11}{3} - \frac{2}{3N}\,.
 \end{equation} 
At weak coupling, $ L \Lambda  \ll 1$,   the bifermion 
condensate in QCD(F)* receives its dominant contribution from the 
instanton-monopole  with the fermion zero modes insertion,
the first term in the second line in Eq.~(\ref{Eq:dQCD(F)}). The condensate 
is proportional to  
\beq
\langle \lambda \psi\rangle\sim
e^{-S_0} \sim e^{- \frac{8 \pi^2 }{g^2N}}\,.
\eeq
Above the  scale $L \Lambda  \sim 1$ 
we expect the bifermion condensate to be $L$-independent and saturate  its value on $R_4$,
   \begin{eqnarray}
\langle \bar \Psi  \Psi \rangle  \sim  \left\{ \begin{array}{ll} 
   \Lambda^3 (\Lambda L)^{b_0-3}  =  \Lambda^3 (\Lambda L )^{(2/3) (1-N^{-1})} \,,&  \quad 
    L \Lambda  \ll 1 \,,  \\[3mm]  
                                   \Lambda^3 \Big (1+  \O (\frac{1}{\Lambda L} ) \Big)\,,  & \quad
                                    L \Lambda \gsim 1\,.
                                            \end{array} \right.
\end{eqnarray}
The above formula is testable on lattices. 

It is natural to believe the saturation scale is associated with the transition from weak to strong coupling and restoration of the spontaneously broken gauge symmetry 
U$(1)^{N-1}\rightarrow {\rm SU}(N)$.  This is the regime where 
the theory passes from the Abelian  to non-Abelian confinement. The effective 
theory (\ref{Eq:dQCD(F)}) which is only valid at $L\Lambda\ll 1$ looses 
its validity when this parameter becomes of order one. Nonetheless, we do 
not expect  phase transitions (or rapid crossovers) 
in the parameter $L \Lambda$. We expect physics of the two regimes to be continuously connected.    

 It would be immensely  useful to study this passage on lattices.   
 In the strong coupling regime, the volume dependent factors enter in
 observables only  via subleading $\O(1/(L  \Lambda))$ terms.   

\section{QCD with one bifundamental  fermion}
\label{s4}

Consider orbifold  QCD, a gauge theory  with 
the SU$(N)_1\times  {\rm SU}(N)_2$   gauge group, and one bifundamental  Dirac fermion, defined on $R_3 \times S_1$,  
\begin{equation}
S^{\rm QCD(BF)}= \int_{R_3 \times S_1} \frac{1}{g^2}\, \Tr \left[ \frac{1}{2} F_{1, MN}^2 (x) + 
\frac{1}{2} F_{2, MN}^2 (x)   +
 i \bar  \Psi  \Dslash  \Psi  \right]\,,
\label{eq:QCDBF}
\end{equation}
where $$ D_M \Psi=  \partial_M \Psi + i A_{1,M} \Psi - i \Psi A_{2, M}\,.$$ 
The theory  possesses  a U(1$)_V \times (Z_{2N})_A \times (Z_2)_I$ symmetry  
which acts on the elementary fields as 
 \begin{eqnarray}
&&U(1)_V: \; \; \;    \lambda \rightarrow e^{i \alpha} \lambda, \;\;  \psi \rightarrow e^{-i \alpha} \psi,  
\nonumber\\[2mm]
&&(Z_{2N})_A: \; \;   \lambda \rightarrow e^{i \frac{2 \pi}{2N}}  \lambda, \;\; 
 \psi \rightarrow e^{i \frac{2 \pi}{2N}} \psi,
  \nonumber\\[2mm]
 &&(Z_2)_I:  \;\; \; \;\;   \lambda \leftrightarrow  \psi, \; \;   A_{\mu, 1} \ \leftrightarrow  A_{\mu, 2}  \,.
 \label{Eq:symorb}
\end{eqnarray}
The  $(Z_{2N})_A$  symmetry is the anomaly-free subgroup of the axial 
U(1)$_A $.  It is a folklore statement that with sufficiently large $r(S_1)$, 
the chiral symmetry is broken down to $Z_2 $ by the formation of the bifermion  condensate, 
\beq
 \langle \bar \Psi \Psi \rangle  = 4N \Lambda^3 \cos\left({\frac{2 \pi k}{N}}  \right)\,,
 \qquad k=0,\,1,\, ...\, N-1\,,
\eeq
marking $N$ isolated vacua in the same manner as in $\N=1$ SYM theory. 

QCD(BF) on $R_4$ is believed confine in the same way as $\N=1$ SYM
theory, possesses a mass gap, and $N$ isolated vacua.  We would like to shed some light on these issues by studying 
QCD(BF)* with small $r(S_1)$.  

\subsection{Deformed orbifold QCD}

On  $S_1 \times R_3$ we can deform original QCD(BF)
\begin{eqnarray}
S&=&  \int_{R_3}  \;  \frac{L}{g^2} \Tr  \Big[ \frac{1}{2} F_{1, \mu \nu}^2 + \frac{1}{2} F_{2, \mu \nu}^2 +  (D_{\mu} \Phi_1)^2 +  (D_{\mu} \Phi_2)^2
 +  g^2  V [\Phi_1,\Phi_2]   \qquad 
 \nonumber\\[3mm]
 && +  i \bar  \lambda 
\Big( \sigma_{\mu} (\partial_{\mu} \lambda + i A_{1, \mu} \lambda - i \lambda  A_{2, \mu})  + i \sigma_4 (\Phi_1 \lambda - \lambda  \Phi_{2})     \Big)  
\nonumber\\[3mm]
 && 
  +  i \bar \psi
\Big( \sigma_{\mu} (\partial_{\mu} \psi - i A_{1, \mu} \psi + i \psi  A_{2, \mu})  - i \sigma_4 
(\Phi_1 \psi - \psi \Phi_{2})     \Big)
    \Big]   \,,
\end{eqnarray}
by adding double-trace terms (\ref{fourma}) in such a way that the center 
symmetry is not broken in the vacuum.   
The  center symmetry stability at weak coupling  implies that the vacuum of the theory is located at 
\beqn
L \langle \Phi_1\rangle = L \langle \Phi_2 \rangle =&& {\rm diag} \left( -\frac{2\pi [N/2]}{N},\,\,  -\frac{2\pi ([N/2]-1)}{N}, ....,\,
\frac{2\pi [N/2]}{N} \right) \,, \nonumber\\[2mm]
&&({\rm mod } \;  2 \pi)\,,
 \label{Eq:pattern}
\eeqn
cf. Eq.~(\ref{12}).  
Consequently, in the weak coupling regime, the gauge symmetry is broken,
\begin{equation}  
[{\rm SU}(N)]_1 \times [{\rm SU}(N)]_2  \longrightarrow [{\rm U}(1)^{N-1}]_1 \times [{\rm U}(1)^{N-1}]_2\,.
\end{equation}

In perturbation theory $2(N-1)$  photons remain massless while
all off-diagonal gauge fields acquire masses in the range $\left[\frac{2 \pi}{L},    
\frac{2 \pi}{L} \right]$. The three-dimensional mass terms 
of the bifundamental fermions are determined by
$$
\sum_{i,k=1}^N\, (a_i^1 -a_k^2) \overline{ \Psi^k_i} \gamma_4 \Psi_i^k
$$
where $a_k^1, a_k^2$ are the eigenvalues of 
$\Phi_1$ and $\Phi_2$, see Eq.~(\ref{Eq:pattern}).
The diagonal components of the bifundamental fermions 
$$
\left(\lambda^i_k\,,\,\,\,\psi^k_i
\right)_{i=k}
$$
remain massless to all orders in perturbation theory; 
we will refer to them as $\lambda_i,\,\,\psi_i$ ($i=1, ..., N$).
Other components get masses $\sim 2\pi  (i-k)/L $, and decouple in the low-energy limit,
and so do the $W$ bosons.

The bifundamental fermions are electrically charged under the unbroken 
$ [{\rm U}(1)^{N-1}]_1 \times [{\rm U}(1)^{N-1}]_2$  in a correlated fashion. 
If in Sect. \ref{FF} the electric charges of each fermion were
characterized by an $(N-1)$-dimensional vector
$\mbox{\boldmath $q$}_{\Psi_i}$, now they are characterized by concatenation of two such 
$N-1$ dimensional electric charge vectors
\beq
\mbox{\boldmath $q$}_{\lambda_i} = g\, ( + \mbox{\boldmath $H$}_{ii},  - \mbox{\boldmath $H$}_{ii}
 ) \,,\quad \mbox{\boldmath $q$}_{\psi_i} = g\, ( - \mbox{\boldmath $H$}_{ii},  + \mbox{\boldmath $H$}_{ii}
 ) \,,\quad
i=1, ... , N\,,
\eeq
Thus, the low-energy effective  Lagrangian in perturbation theory is 
 \begin{eqnarray}
 &&
S^{\rm pert\,\, th}=  \int_{R_3}  \;  \frac{1}{g_3^2}\,  \Big[ \sum_{a=1}^{N-1}\,
\Big( \frac{1}{4} F^{a, 2}_{1, \mu \nu} + \frac{1}{4} F^{a, 2}_{2, \mu \nu}  \Big) 
\nonumber\\[3mm]
 &&+ 
    \sum_{i=1}^{N}   i \bar \Psi_i \gamma_{\mu} \Big( \partial_{\mu} + i
     \mbox{\boldmath $H$}_{ii} \mbox{\boldmath $A$}_{\mu}^{1}  - i
      \mbox{\boldmath $H$}_{ii} \mbox{\boldmath $A$}_{\mu}^{2}    
 \Big)  
      \Psi_i  
    \Big]   \,.
\end{eqnarray}
The mass gap must arise due to nonperturbative effects, as in Sect.~\ref{FF}. We 
will identify and classify nonperturbative effects induced by topologically
nontrivial field configurations momentarily. 

\subsection{Nonperturbative low-energy effective Lagrangian}

Nonperturbatively, the gauge symmetry breaking pattern (\ref{Eq:pattern}) implies the 
existence of  $N$ types  of  instantons-monopoles associated  with each gauge group. 
The magnetic and topological charges of these  objects are 
\begin{eqnarray}
\left( \int_1 F,  \int_1  \frac{g^2}{32 \pi^2}  \, F^a {\widetilde F}^a\,; \,   \int_2 F, \int_2  \frac{g^2}{32 \pi^2}  \, F^a {\widetilde F}^{a}   \right) =   
\left\{ \begin{array}{ll}
\left(  \pm  \frac{4 \pi}{g} 
\mbox{\boldmath $\alpha$}_i\,, \pm \frac{1}{N}, 0, 0\right) \,,\\[4mm]
\left( 0, 0,  \pm  \frac{4 \pi}{g} 
\mbox{\boldmath $\alpha$}_i\,, \pm \frac{1}{N}\right) \,.
\end{array}
\right.
\end{eqnarray}
Consequently,  each monopole generates two fermion zero modes, and 
the instanton-monopole vertices are 
\begin{eqnarray}
&& 
{\cal M}_{i}^{1} : ( + \frac{4\pi}{g} \mbox{\boldmath $\alpha$}_i\,, + \frac{1}{N}, 0, 0) : \;\;  e^{+i \mbox{\boldmath $\alpha$}_i\,  \mbox{\boldmath $\sigma$}_1 } ({ \lambda_i  \psi_i +
 \lambda_{i+1}  \psi_{i+1}   } )\,, \nonumber\\[3mm]
&&\overline {\cal M}_{i}^{1} : (  -\frac{4\pi}{g} \mbox{\boldmath $\alpha$}_i\,,  - \frac{1}{N}, 0, 0) :\; \;\,  e^{-i \mbox{\boldmath $\alpha$}_i\, \mbox{\boldmath $\sigma$}_1 } 
({\bar \lambda_i \bar \psi_i +
 \bar \lambda_{i+1}   \bar \psi_{i+1}    } ) \,,
\nonumber\\[3mm]
&&{\cal M}_{i}^{2} : ( 0, 0, +\frac{4\pi}{g}  \mbox{\boldmath $\alpha$}_i\,, + \frac{1}{N} ) :\; \;\,\,  e^{+i \mbox{\boldmath $\alpha$}_i\, \mbox{\boldmath $\sigma$}_2 }
 ({ \lambda_{i} 
  \psi_{i} +  \lambda_{i+1}  \psi_{i+1}    } )\,,
\nonumber\\[3mm]
&&\overline {\cal M}_{i}^{2}:   ( 0, 0,  -\frac{4\pi}{g} \mbox{\boldmath $\alpha$}_i\,,  - \frac{1}{N}) :\; \;\,\,
 e^{- i \mbox{\boldmath $\alpha$}_i\, \mbox{\boldmath $\sigma$}_2 } 
({ \bar \lambda_{i}   \bar \psi_{i}  +
 \bar \lambda_{i+1}  \bar \psi_{i+1}    } ) \,,
\end{eqnarray}
where $\mbox{\boldmath $\sigma$}_1$ is the set of dual photons for
$[{\rm U}(1)^{N-1}]_1$ while $\mbox{\boldmath $\sigma$}_2$
is the set of dual photons for
$[{\rm U}(1)^{N-1}]_2$.
In full analogy with 
the SYM theory,  the $2N$ fermion zero modes of the BPST instanton 
split  into $N$ pairs: each instanton-monopole   supports two fermion zero modes. This is a natural consequence of the Callias index theorem. (The same conclusion was also reached by Tong  \cite{Tong:2002vp}).    

As a result, the  instanton-monopole contributions give rise to
the  following  terms in the  effective Lagrangian: 
   \beqn
   && \Delta L^{\rm QCD(BF)*}=
   {\rm const.}  \times\;   g^{-6}  e^{ -S_{0}} 
    \sum_{\alpha_{i} \in \Delta_{\rm aff}^{0}}
    \Big(  \left(e^{i \mbox{\boldmath $\alpha$}_i\, \mbox{\boldmath $\sigma$}_1 } +  e^{i \mbox{\boldmath $\alpha$}_i\, \mbox{\boldmath $\sigma$}_2 }\right)   
    \nonumber\\[3mm]
    &&\times
    ( \lambda_i  \psi_i +  \lambda_{i+1}  \psi_{i+1}    )
      +  {\rm H.c.}
                  \Big)\,.
                  \label{51}
     \eeqn
At the level $e^{-S_0}$  the 
instanton-monopole effects in QCD(BF)*  cannot  provide mass terms for the dual photons. This situation is completely analogous  to that in QCD(Adj)* where all 
instanton-monopoles have fermion zero modes  and, hence, are unable to contribute to the bosonic potential for the dual photons \mbox{\boldmath $\sigma$}$_1$
 and \mbox{\boldmath $\sigma$}$_2$. 

The situation drastically changes at order $e^{-2S_0}$. There are  nontrivial effects which render the long-distance three-dimensional
fields  massive, implying confinement. An easy way to see that this is the case
is to examine the symmetries of the theory.  

Since  U(1)$_V \times (Z_{2N})_A \times (Z_2)_I$ is the symmetry 
of the microscopic theory, 
it must be manifest in  the  low-energy effective theory in three dimensions.  
The invariance of the instanton-monopole vertex  under U(1)$_V$ and   $(Z_2)_I$ is manifest.  At the same time, the $(Z_{2N})_A$
invariance requires combining the axial chiral symmetry 
with the  discrete shift symmetry of the dual photon, 
 \beqn
(Z_{2N})_A: \; \;\;\;  &&\lambda \psi \rightarrow e^{i \frac{2 \pi}{N}} \lambda \psi, \nonumber\\[3mm]
 &&
 \mbox{\boldmath $\sigma$}_{1,2}   \rightarrow \mbox{\boldmath $\sigma$}_{1,2}
 - \frac{2 \pi}{N} \mbox{\boldmath $\rho$}  \qquad 
 \label{Eq:symorb2}
\eeqn
where $ \mbox{\boldmath $\rho$} $ is the Weyl vector defined by 
\beq
\mbox{\boldmath $\rho$} =  \sum_{j=1}^{N-1} \mbox{\boldmath $\mu$}_k\,, 
\label{dop1}
\eeq
and \mbox{\boldmath $\mu$}$_k$ stand for the $N-1$ fundamental weights 
of the associated Lie algebra, defined through the  reciprocity relation, 
\beq
\frac{2 \mbox{\boldmath $\alpha$}_i \mbox{\boldmath $\mu$}_j }
{ \mbox{\boldmath $\alpha$}_i^{2}}=   \mbox{\boldmath $\alpha$}_i \mbox{\boldmath $\mu$}_j     = \delta_{ij}\,.
\label{dop2}
\eeq
Using the identities 
\begin{equation}
 \mbox{\boldmath $\alpha$}_N   \mbox{\boldmath $\rho$} =  -(N-1) \,,  \quad \mbox{\boldmath $\alpha$}_i  \mbox{\boldmath $\rho$}= 1\,
  , \quad  i=1,\,  \ldots\,  N-1\; , 
\label{iden}
\end{equation}
 the vertex  operator  
\begin{equation}
e^{i \mbox{\boldmath $\alpha$}_i\,  \mbox{\boldmath $\sigma$}_{1,2} } 
\rightarrow e^{i \mbox{\boldmath $\alpha$}_i\,
 ( \mbox{\boldmath $\sigma$}_{1,2} -\frac{2 \pi}{N}   
\mbox{\boldmath $\rho$}) } =  
e^{-i \frac{2 \pi}{N} } \;
e^{i \mbox{\boldmath $\alpha$}_i\,  \mbox{\boldmath $\sigma$}_{1,2}} 
 \,, \quad i=1,\, \ldots, \, N\,,
\end{equation}
rotates in the opposite direction compared with the fermion bilinear,
by the same amount. Hence, the instanton-monopole induced    vertex    
 $$(e^{i \mbox{\boldmath $\alpha$}_i\, \mbox{\boldmath $\sigma$}_1 } +  e^{i \mbox{\boldmath $\alpha$}_i\, \mbox{\boldmath $\sigma$}_2 })  
      ( \lambda_i  \psi_i +  \lambda_{i+1}  \psi_{i+1} )      $$ 
is invariant under  the discrete chiral symmetry.  

The discrete shift symmetry, (\ref{Eq:symorb2}) as opposed to the continuous shift symmetry, cannot prohibit  mass term for the dual photons. At best, it can 
postpone its  appearance  in the $e^{-S_0}$ expansion.  Hence, such 
a mass term must be, and is, generated. 

As in  SYM theory,  at level $e^{-2S_0}$ 
there exist magnetically charged   bound  monopole-antimonopole pairs 
with no fermion zero modes.  These stable pairs were referred to as magnetic bions in 
\cite{Unsal:2007jx}. 
In QCD(BF)*, the bions come in a wider variety than in
SYM theory. The analogs of the magnetic  bions that appear in SYM 
theory are the pairs of the type
${\cal M}_i^{1}$ and $\overline {\cal M}_{i\pm1} ^{1}$ (and  $1 \leftrightarrow 2$).  
Despite the repulsive Coulomb interactions between these two monopoles they form bound states due to the fermion exchange between them, with the  combined effect
$$
\sim \frac{1}{r} + \log r\,.
$$
The corresponding bound state is stable. 

Since the fermion zero modes  in QCD(BF)* communicate with the mono\-poles 
in both  gauge groups, the fermion zero mode exchange also  generates logarithmic attractive interactions between the monopoles 
${\cal M}_i^{1}$  in the first gauge group and the antimonopoles 
$\overline {\cal M}_{i, i\pm1} ^{2}$ in the second. Note that  there is
no Coulomb interaction between  these two since the first 
instanton-monopole is charged under the  [U(1)  $^{N-1}]_1 $  gauge subgroup of  [U(1)$^{N-1}]_1 \times [{\rm U}(1)^{N-1}]_2 $ while the second is charged under [U(1)$^{N-1}]_2$. 
Thus, the stable magnetic bions in QCD(BF)*, their magnetic and topological charges,
and the vertices they generate are 
 \begin{eqnarray}
&&{\cal B}^1_i : \Big(   \frac{4\pi}{g} (\mbox{\boldmath $\alpha$}_i\, - \mbox{\boldmath $\alpha$}_{i-1} ), \; 0, 0, 0 \Big)  :  \qquad  c_1 e^{-2S_0} e^{i ( \mbox{\boldmath $\alpha$}_i\, - \mbox{\boldmath $\alpha$}_{i-1}) \mbox{\boldmath $\sigma$}_1} \nonumber\\[3mm]
&&{\cal B}^2_i:  \Big(0, 0,   \frac{4 \pi}{g} (\mbox{\boldmath $\alpha$}_i\, - \mbox{\boldmath $\alpha$}_{i-1}), 0 \Big )  :  \qquad c_1 e^{-2S_0} e^{i ( \mbox{\boldmath $\alpha$}_i\, - \mbox{\boldmath $\alpha$}_{i-1}) \mbox{\boldmath $\sigma$}_2}  \nonumber\\[3mm]
& & {\cal B}^{12}_{i,i} :  \Big(   \frac{4\pi}{g} \mbox{\boldmath $\alpha$}_i\,, \frac{1}{N},   -    \frac{4\pi}{g} \mbox{\boldmath $\alpha$}_{i}, - \frac{1}{N} \Big)  : \qquad 
  c_2 e^{-2S_0} e^{i (   \mbox{\boldmath $\alpha$}_i\,   
  \mbox{\boldmath $\sigma$}_1 -
  \mbox{\boldmath $\alpha$}_{i } \mbox{\boldmath $\sigma$}_2) }  \nonumber\\[3mm]
  & & {\cal B}^{12}_{i,i-1} : \Big (  \frac{4\pi}{g}  \mbox{\boldmath $\alpha$}_i\,,  \frac{1}{N},  -    \frac{4\pi}{g} \mbox{\boldmath $\alpha$}_{i-1},  - \frac{1}{N} \Big)  :  c_2 e^{-2S_0} e^{i ( \mbox{\boldmath $\alpha$}_i\, \mbox{\boldmath $\sigma$}_1 -
\mbox{\boldmath $\alpha$}_{i-1 }\mbox{\boldmath $\sigma$}_2) }  \nonumber\\[3mm]
& & {\cal B}^{12}_{i, i+1} : \Big  (   \frac{4\pi}{g} \mbox{\boldmath $\alpha$}_i\,, \frac{1}{N},   -    \frac{4\pi}{g} \mbox{\boldmath $\alpha$}_{i+1}, - \frac{1}{N}\Big )  : c_2  e^{-2S_0} e^{i ( \mbox{\boldmath $\alpha$}_i\, \mbox{\boldmath $\sigma$}_1 -
  \mbox{\boldmath $\alpha$}_{i+1 } \mbox{\boldmath $\sigma$}_2) } 
\end{eqnarray}
The vertices  for antibions (such as $\overline {\cal B}^1_i$)   are the complex conjugates of the ones given above.   The above bions are stable due to the 
 attractive fermion pair exchange between their constituents. 
Note that the constituents of the bions 
${\cal B}^1_i $  and  ${\cal B}^2_i $, unlike the ones of  
  $ {\cal B}^{12}_{i,i} ,  {\cal B}^{12}_{i,i+1} ,  {\cal B}^{12}_{i,i-1}$
need to  compete with  the Coulomb repulsion for stability. 
Thus, in principle, 
there are no (symmetry or microscopic) reasons  for the prefactor of the first two 
 to be the equal to the ones of the latter.  Therefore, we assume they are not. 

As a result,  we obtain the bion-induced bosonic potential in  QCD(BF)* in the form
  \begin{eqnarray}
  &&  V_{\rm bion} ( \mbox{\boldmath $\sigma$}_1, \mbox{\boldmath $\sigma$}_2 ) = 
  m_W^3 g^{-6}
  e^{-2S_0}    \sum_{i=1}^{N} \Big[ 
     c_1 \Big( 
     e^{i ( \mbox{\boldmath $\alpha$}_i \, - \mbox{\boldmath $\alpha$}_{i-1} \,) \mbox{\boldmath $\sigma$}_1} + e^{i ( \mbox{\boldmath $\alpha$}_i
     \, - \mbox{\boldmath $\alpha$}_{i-1} \,) \mbox{\boldmath $\sigma$}_2}
     \Big)      
     \nonumber\\[3mm]     
&&  +    c_2 \Big( 
2 e^{i ( \mbox{\boldmath $\alpha$}_i\, \mbox{\boldmath $\sigma$}_1 -
 \mbox{\boldmath $\alpha$}_i\,  \mbox{\boldmath $\sigma$}_2) } + 
  e^{i ( \mbox{\boldmath $\alpha$}_i\, \mbox{\boldmath $\sigma$}_1 -
  \mbox{\boldmath $\alpha$}_{i-1}\,  \mbox{\boldmath $\sigma$}_2) }
  + e^{i ( \mbox{\boldmath $\alpha$}_i \, \mbox{\boldmath $\sigma$}_1 -
\mbox{\boldmath $\alpha$}_{i+1}\,   \mbox{\boldmath $\sigma$}_2) }   
  \Big)             
       \Big]          
       +  {\rm  H.c.} \nonumber
       \\    
           \end{eqnarray}
In full analogy with the superpotential in SYM* theory, 
it is convenient to
define a prepotential in QCD(BF)*.  To this end we  introduce the function
\begin{equation}
{\mathcal W}( \mbox{\boldmath $\sigma$}_1, 
\mbox{\boldmath $\sigma$}_2) =  m_W g^{-4} e^{-S_0}  \sum_{ \mbox{\boldmath $\alpha$}_i \, \in \Delta_0^{\rm aff}}   \left( e^{i \mbox{\boldmath $\alpha$}_i\, \mbox{\boldmath $\sigma$}_1 } 
+ e^{i \mbox{\boldmath $\alpha$}_i \, \mbox{\boldmath $\sigma$}_2 }  \right) \,,
\label{Eq:prepotential}
\end{equation}
to be referred to as prepotential.
Note that the prepotential, as well as its derivatives,  transform homogeneously under the 
$Z_{2N}$ shift symmetry (\ref{Eq:symorb2}),       
$$
Z_{2N}: \quad {\mathcal W}( \mbox{\boldmath $\sigma$}_1, \mbox{\boldmath 
$\sigma$}_2)      
 \longrightarrow     e^{-i \frac{2 \pi}{N}}       {\cal W}( \mbox{\boldmath $\sigma$}_1, \mbox{\boldmath $\sigma$}_2)  \,.
$$
Now, it is easy to express the bion-induced potential in terms of the
prepotential in the form which is 
manifestly  invariant under the $Z_{2N}$ shift  and  $(Z_2)_I$ interchange symmetries,
 \begin{equation}
V( \mbox{\boldmath $\sigma$}_1, \mbox{\boldmath $\sigma$}_2) = g_3^2 \sum_{a=1}^{N-1} \left(\;  c_+ \left |\frac{\partial {\cal W}}{\partial \sigma_{1,a}}  + 
\frac{\partial {\cal W}}{\partial \sigma_{2,a}}   
\right|^2  +  c_{-} \left|\frac{\partial {\cal W}}{\partial \sigma_{1,a}}  - 
\frac{\partial {\cal W}}{\partial \sigma_{2,a}}   
\right|^2  \; \right)\,.
\label{Eq:potentialBF}
\end{equation}      
We are finally ready to present the low-energy effective theory for QCD(BF)*,
\begin{eqnarray}
&&L^{\rm QCD(BF)^*} = 
 \frac{g_3^2}{32 \pi^2} \left[ (\partial  \mbox{\boldmath $\sigma$}_1)^2
   +  (\partial  \mbox{\boldmath $\sigma$}_2)^2
   \right] + 
V_{\rm bion} ( \mbox{\boldmath $\sigma$}_1, \mbox{\boldmath $\sigma$}_2 )       
       \nonumber\\[4mm]
 &&  +   \frac{1}{g_3^2} \sum_{i=1}^{N}   i  \overline \Psi_i \gamma_{\mu} \Big(  \partial_{\mu} + 
 i \mbox{\boldmath $H$}_{ii} \mbox{\boldmath $A$}_{\mu}^{1}  - 
     i  \mbox{\boldmath $H$}_{ii} \mbox{\boldmath $A$}_{\mu}^{2}    
  \Big)  
      \Psi_i   
   \nonumber\\[4mm]
&&  
 +   c \; g^{-6}  e^{ -S_{0}} 
    \sum_{\alpha_{i} \in \Delta_{\rm aff}^{0}}
    \Big(  
     ( e^{+i \mbox{\boldmath $\alpha$}_i\,  \mbox{\boldmath $\sigma$}_1 }  + 
     e^{+i \mbox{\boldmath $\alpha$}_i\,  \mbox{\boldmath $\sigma$}_2 }) 
     ({ \lambda_i  \psi_i +
 \lambda_{i+1}  \psi_{i+1}   } )\, + 
          { \rm H.c.}           
        \Big) \,.
           \nonumber\\
\label{Eq:dQCDBF2} 
\end{eqnarray} 

Like in other QCD-like theories with complex-representation fermions (such as 
QCD(F/AS/S)*), but unlike the ones with real-representation fer\-mions (such as
SYM theory or QCD(adj)), we have both the electric and magnetic couplings. 
The Lagrangian (\ref{Eq:dQCDBF2}) includes all relevant terms allowed by symmetries up to $\O(e^{-3S_0})$. 
 
The important question at this stage is  which operators 
in our effective Lagrangian (\ref{Eq:dQCD(F)}) are most important  at large distances  in the renormalization-group sense. Apparently, the fugacity (the coefficient in front  of the bion  vertices) has dimension $+3$ and is dominant in the IR. The quantum-mechanical corrections are negligible. This suggests that 
in the IR the effects produced by magnetically charged bions are most relevant. 
 
 \subsection{Vacuum structure and chiral symmetry realization}
 
The low-energy effective theory respects all symmetries of the underlying 
gauge theory ${\rm U}(1)_V \times (Z_{2N})_A \times (Z_2)_I$ and  $C, P, T$.  
These symmetries may be  spontaneously broken. By studying dynamics of the effective theory we demonstrate that the breaking pattern is 
\begin{eqnarray}
{\rm U}(1)_V \times (Z_{2N})_A \times (Z_2)_I \rightarrow {\rm U}(1)_V \times (Z_{2})_A \times (Z_2)_I 
\end{eqnarray}
leading to the occurrence of $N$ isolated vacua.  
 
In  Eq.~(\ref{Eq:dQCDBF2}) the $Z_{2N}$ chiral symmetry is entangled with the shift symmetry of the dual photon (\ref{Eq:symorb2}), just like in SYM theory.  There are $N$ isolated vacua in the $(Z_2)_I$ invariant subspace related to each other by the 
action of the $Z_N$ shift symmetry. These vacua are located at 
 \begin{equation}
 \mbox{\boldmath $\sigma$}_1= \mbox{\boldmath $\sigma$}_2 = \left\{ 0, \,\frac{2 \pi}{N}, \, \frac{4 \pi}{N},\, \ldots,  \, \frac{2 (N-1)\pi}{N} \right\}  \mbox{\boldmath $\rho$} 
 \end{equation}
in the field space.  The choice of a given vacuum spontaneously breaks 
the $Z_{N}$ shift symmetry, and, hence, the chiral symmetry.
  
Let $|\Omega_k\rangle$ denote one of the $N$ vacuum states ($k=1,\, \ldots ,\, N$). Following the techniques of \cite{Davies:1999uw, Davies:2000nw},  
we observe that the chiral condensate  is proportional to 
the monopole-induced term $e^{-S_0}$.  The renormalization-group $\beta$ function of QCD(BF)* is identical to that of SYM theory up to $\O(1/N^2)$ corrections. The first coefficients are just identical. Thus, 
\beq
 e^{-S_0} \equiv e^{-\frac{8 \pi^2}{g^2N}} =   \Lambda^3 (\Lambda  L)^{b_0-3} 
 \label{eso}
\eeq
where  $b_0$ denotes  the leading coefficient of the $\beta$ function divided by $N$. 
At one-loop order in QCD(BF)*  
$$b_0=  3\,.$$ 
Thus, the chiral condensate in QCD(BF)* is  
    \begin{eqnarray}
\langle \Omega_k| \tr \bar \Psi   \Psi | \Omega_k \rangle  =    2 N  \Lambda^3 
 e^{i \frac { 2 \pi k}{N}}+ {\rm H.c.}\,.
\label{Eq:condensate}
\end{eqnarray}
There is no $L$ dependence in the condensate in QCD(BF)* at one-loop level, just like in
SYM theory. 

\subsection{Mass gap and confinement}  
\label{sec:mas}

The small fluctuation analysis around any of the $N$ minima  is sufficient to see that there are no massless modes in the infrared description of the QCD(BF)*.  The choice of the vacuum breaks the discrete chiral symmetry rendering all fermions massive.  
The bion-induced potential makes all $2(N-1)$ photons massive. This shows that every particle-like excitation must have a finite mass $m \sim e^{-S_0}$. There are no physical states in the mass range  $[0, m) $ in the physical Hilbert space of the theory.  Since the global $Z_N$ center  group symmetry and $(Z_2)_I$ interchange symmetry are unbroken, the physical states can be expressed as the mutual  eigenstates of these symmetries. The Fourier transform  
\begin{equation}
{\sigma}_{\pm, k} = ( \sigma_{1,k} \pm  \sigma_{2,k})
\, \equiv \,
 \frac{1}{\sqrt N} \sum_{j=1}^{N} e^{i \frac{2 \pi j k}{N}}   \mbox{\boldmath $H$}_{jj}  (\mbox{\boldmath $\sigma$}_{1} \pm   \mbox{\boldmath $\sigma$}_{2})
\end{equation}
diagonalizes the mass matrix. 
The masses of the dual photons are proportional to $\exp (-S_0)$. 
More 
exactly,\footnote{Powers of $g$ and numerical factors are omitted
here and in similar expressions below.}
\begin{eqnarray}
 m_{\sigma_{\pm,k}} = \sqrt {c_{\pm}}\,\, \Lambda (\Lambda  L )^2  \left(2 \sin \frac{\pi k }{N}\right)^2,  
 \qquad \Lambda L \ll 1\,.
\end{eqnarray}
Any probe charge one might consider is coupled to a number of $\sigma$
fields.  The thickness of the
domain line (string)
attached to the probe charge
is determined by the inverse mass of the lightest $\sigma$ field ($k=1$).
It is worth noting that the string has a substructure corresponding
to the contribution of the next-to-lightest, next-to-next-to-lightest
and so on $\sigma$'s.
 The fermion masses are of the same order of magnitude in the same regime, as seen from Eq.~(\ref{51}),
\beq
 m_{\Psi_{i}} =  c \Lambda (\Lambda  L )^2\,.
 \eeq

Now we are ready to discuss strings in QCD(BF)* at small $L$. 
Let us consider a heavy probe quark $Q^{i_1 ... i_m}_{j_1 ... j_n}$ and its antiquark 
$\overline{Q^{i_1 ... i_m}_{j_1 ... j_n}}$ in a color-singlet state
at an exponentially large distance from each other. If $m\neq n$ the string (domain line)
forming between these probe objects is unbreakable.
Light dynamical fermions of the low-energy theory
cannot screen the electric charges of the probe quarks.
However,
if $m=n$ some strings (i.e. those attached to the
probes for which every index $i$ is equal to some $j$)
will break through pair creation of light dynamical fermions.
Assume $|n-m|\equiv k \neq 0$. Then the tensions of these unbreakable
$k$ strings can be found by calculating the tensions of the domain lines
supported by the theory (\ref{Eq:dQCDBF2}). These tensions are of the order of $
\Lambda^2 (\Lambda L)$ in the $\Lambda L \ll 1$ Abelian confinement regime 
while at $\Lambda L \gsim  1$, in the non-Abelian confinement  regime,
they tend to $\Lambda^2$ times a numerical coefficient. 

To the best of our knowledge, this is the first analytic demonstration of  $\chi$SB, mass gap generation and linear confinement in QCD(BF)*. This theory  exhibits all expected nontrivial features of QCD(BF) on $R_4$.

\section{QCD with one AS fermion}
\label{s5}

Now we will discuss QCD with one antisymmetric Dirac fermion\,\footnote{Discussion of QCD with the symmetric representation fermion is parallel.}   on $R_3 \times S_1$.
The theory possesses a U(1)$_V \times Z_{2N-4} $ symmetry, $Z_{2N-4} $
being the anomaly-free subgroup of the axial  U(1)$_A $. 
The action of the symmetry on  the elementary fields is as follows:
 \begin{eqnarray}
&&
U(1)_V: \; \; \;\;\;\;\;\; \;  \lambda \rightarrow e^{i \alpha} \lambda, \;\; \qquad 
 \psi \rightarrow e^{-i \alpha} \psi\,,  \nonumber\\[3mm]
&&
(Z_{2N-4})_A: \; \;\;\;   \lambda \rightarrow e^{i \frac{2 \pi}{2N-4}}     \lambda\,,
\;\; \qquad 
 \psi \rightarrow e^{i \frac{2 \pi}{2N-4}} \psi\, .
  \label{Eq:symori}
\end{eqnarray}
It is believed that for sufficiently large  $r(S_1)$,  
the chiral symmetry is broken down to $Z_2 $ by the  
bifermion  condensate  $\langle \psi \lambda \rangle\neq 0$,
$$
 \langle \bar \Psi  \Psi \rangle  \sim N \Lambda^3 e^{i\frac{2 \pi k}{N-2}}  
 +{\rm H.c.}
$$
resulting in $N-2$ isolated vacua . 
The QCD(AS) theory on $R_4$ must confine the same way as $\N=1$ SYM
theory and  possess a mass gap. Since the discussion is quite similar to the 
case of QCD(BF)*, we will be brief. 
 
\subsection{Deformed orientifold QCD}

In the small $r(S_1)$ regime,  the gauge symmetry is broken, 
SU$(N) \rightarrow {\rm U}(1)^{N-1} $.  
Without loss of generality we can take $N=2m+1$.
The case $N=2m$ can be dealt with in a similar manner.  

In perturbation theory the massless fields are
$N-1$ diagonal photons and  $N-2$ charged fermions. The  $N^2-N$ 
off-diagonal $W$ bosons and  
$N^2 - 2N +2$  fermions acquire masses in the range $[\frac{2\pi}{LN}, \frac{2\pi}{L}) $ and decouple from infrared physics.  

The AS fermions $\Psi_{ij}$ acquire three-dimensional mass terms  given by  
$$
\sum_{i,j=1}^N\, (a_i  + a_j) \bar \Psi^{[ij]} \gamma_4 \Psi_{[ij]}
$$
where $a_k$'s are given in Eq,~(\ref{12}). Hence, 
$$
m_{ij}= \frac{2 \pi}{LN} \,\big( [i+j ]\; \, {\rm mod}\,\,\, N\big)
\,.
$$
Thus,   
the fermion components  $\Psi_{i, N-i}$ remain massless to all orders in perturbation theory.  Let us label   
$$
\Psi_{i, N-i} \equiv \Psi_{i}\,,\quad  i=1, \ldots , N-1\,.
$$
The electric charges of these 
degrees of freedom under the unbroken gauge group is 
\beq
\mbox{\boldmath $q$}_{\Psi_i} = g\, (  \mbox{\boldmath $H$}_{ii} + 
 \mbox{\boldmath $H$}_{N-i, N-i} ) \,,\quad 
\quad i=1, ... , N\,,
\eeq
Since the fermion is antisymmetric in its indices, we may parameterize the set
of the massless fermions as 
 \begin{eqnarray}  
  \Psi = && \left\{ \Psi_1, \ldots, \Psi_{m-1}, \Psi_{m}, \; \;  \Psi_{m+1}, \Psi_{m+2}, \ldots, \;\; 
    \Psi_{2m}\right\}
   \nonumber\\[2mm]
   =&&  
   \left\{ \Psi_1, \ldots, \Psi_{m-1}, \Psi_{m}, -\Psi_{m}, -\Psi_{m-1}, \ldots ,  - \Psi_{1}
\right\}
 \,.
\end{eqnarray}  
The IR action in perturbation theory is 
 \begin{eqnarray}
S=  \int_{R_3}  \;  \frac{1}{g_3^2} \Big[ \frac{1}{4} \sum_{a=1}^{N-1}  (F^{a}_{\mu \nu})^2   
+  2 \sum_{i=1}^{m}     i \bar \Psi_i \gamma_{\mu} \Big(  \partial_{\mu} + 
 i (  \mbox{\boldmath $H$}_{ii} + \mbox{\boldmath $H$}_{N-i,N-i} ) 
     \mbox{\boldmath $A$}_{\mu} 
 \Big)  
      \Psi_i  
    \Big]   .
    \nonumber\\
\end{eqnarray}

\subsection{Nonperturbative effects}

In  QCD(AS)* on small $S_1 \times R_3$  there are $N$ types of 
instanton-monopoles because of the pattern of the
gauge symmetry breaking SU$(N) \rightarrow {\rm U}(1)^{N-1} $ via a 
compact adjoint Higgs field. 
The $2N-4$ fermion  zero modes  of the BPST $R_4$ instanton split
into $N-2$ pairs of the instanton-monopole zero modes 
in a slightly different way than that in SYM* theory and QCD(BF)*. The $N-2$
instanton-monopoles have 
two fermion zero modes each,  while the remaining two monopoles have no zero 
modes. It is useful to present the monopole-instanton vertices in QCD(AS)* due to
a nontrivial structure of their zero modes,
      \begin{eqnarray}
&& {\cal M}_{1}=   e^{ -S_{0}}  e^{i    \mbox{\boldmath $\alpha$} _1   \mbox{\boldmath 
$\sigma$} } \;  
( \lambda_1  \psi_1 + \lambda_2 \psi_2)   \,,\nonumber\\[3mm]
&& {\cal M}_{2}=   e^{ -S_{0}}  e^{i   \mbox{\boldmath $\alpha$}_2    \mbox{\boldmath 
$\sigma$}  } \;  ( \lambda_2  \psi_2 + \lambda_3 \psi_3)   \,,\nonumber\\[3mm]
&&\ldots \,,\nonumber\\[3mm]
&&{\cal M}_{m-1}=   e^{ -S_{0}}  e^{i  \mbox{\boldmath $\alpha_{m-1}$}    \mbox{\boldmath 
$\sigma$}   } \;  ( \lambda_{m-1}  \psi_{m-1}  + \lambda_{m}  \psi_{m}  ) 
\,,\nonumber\\[3mm]
&& {\cal M}_{m}=  e^{ -S_{0}}  e^{i  \mbox{\boldmath $\alpha$}_{m}    \mbox{\boldmath 
$\sigma$}   } \;
   ( 2 \lambda_{m}  \psi_{m} )
\,,\nonumber\\[3mm]
&&  {\cal M}_{m+1}=  
 e^{ -S_{0}}  e^{i   \mbox{\boldmath $\alpha$}_{m+1}    \mbox{\boldmath 
$\sigma$}   } \;
  ( \lambda_{m}  \psi_{m}  + \lambda_{m-1}  \psi_{m-1}  )  
\,,\nonumber\\[3mm]
 &&\ldots  \,,\nonumber\\[3mm]
 && {\cal M}_{2m-2}=   e^{ -S_{0}}  e^{i   \mbox{\boldmath $\alpha$}_{2m-2}    \mbox{\boldmath 
$\sigma$}  } \;  
 ( \lambda_3  \psi_3 + \lambda_2 \psi_2)    
\,,\nonumber\\[3mm]
&& {\cal M}_{2m-1}=   e^{ -S_{0}}  e^{i  \mbox{\boldmath $\alpha$}_{2m-1}    \mbox{\boldmath 
$\sigma$}   } \;  
   ( \lambda_2  \psi_2 + \lambda_1 \psi_1)   
\,,\nonumber\\[3mm]
   && {\cal M}_{2m}=   e^{ -S_{0}}  e^{i   \mbox{\boldmath $\alpha$}_{2m}    \mbox{\boldmath 
$\sigma$}   } \; \,,
   \nonumber\\[3mm] 
    && {\cal M}_{2m +1}=   e^{ -S_{0}}  e^{i   \mbox{\boldmath $\alpha$}_{2m+1}    \mbox{\boldmath 
$\sigma$}   } \,.
           \end{eqnarray}

Consequently, the contribution to  the QCD(AS)*  Lagrangian induced by 
monopole-instantons takes the form 
   \begin{equation}
\Delta  L  \sim     \sum_{i=1}^{2m+1} \left( {\cal M}_{i }  + \overline  {\cal M}_{i }  
\right) .
  \label{Eq:orientivertex}
     \end{equation}         
Since  $N-2$ the monopoles carry compulsory fermionic       
zero mode insertions, they can not  induce  a mass term for all the dual photons 
if  $N \geq 4$.   
As seen from Eq.~(\ref{Eq:orientivertex}), two of the monopole-instantons do contribute 
to the bosonic potential, but this is insufficient to render all photons massive for 
$N \geq 4$.  (At $N=3$, QCD(AS)* and QCD(F)* are the same theories.)  
Thus, in order to  render all the photons massive, we  need to incorporate 
effects of order $e^{-2S_0}$, and introduce the magnetic bions.  
Before doing so let us show that the underlying symmetries of QCD(AS)* allow mass terms
for all dual photons to be generated. 

Since  U(1)$_V \times (Z_{2N-4})_A$ is the symmetry of the microscopic theory, 
it must be a symmetry of the  long distance theory.  
The invariance under $U(1)_V$ is manifest.  
The invariance under the $(Z_{2N-4})_A$ necessitates intertwining the axial chiral symmetry 
with a   discrete shift symmetry of the dual photon,
 \begin{eqnarray}
(Z_{2N-4})_A: \; \;\;\;  &&
\lambda \psi \rightarrow e^{i \frac{2 \pi}{N-2}} \lambda \psi
 \,,\nonumber\\[3mm]
&&  \mbox{\boldmath $\sigma$}    \rightarrow  \mbox{\boldmath $\sigma$}   
- \frac{2 \pi}{N-2}   \mbox{\boldmath $\rho$}_{AS} \,,
 \label{Eq:AS}
\end{eqnarray}
where  
\beq
 \mbox{\boldmath $\rho$}_{AS}\equiv  \sum_{j=1}^{N-2} 
 \mbox{\boldmath $ \mu $}_{k}
\label{dop3}
\eeq
and $\mu_{k}$ are the $N-1$ fundamental weights 
of the associated Lie algebra. Note that the parameter $\mbox{\boldmath $\rho$}_{AS}$ is not exactly the Weyl vector, which appears 
in SYM* theory  and QCD(BF)*. Rather, it can be represented as
\begin{equation}
 \mbox{\boldmath $\rho$}_{AS} =  \mbox{\boldmath $\rho$} -  \mbox{\boldmath $  \mu$}_{N-1} \; .
\end{equation}
Using the identities 
\begin{equation}
 \mbox{\boldmath $\alpha$}_{N-1} \mbox{\boldmath $\rho$}_{AS} = 0\,, \quad  \mbox{\boldmath $\alpha$}_N   \mbox{\boldmath $\rho$}_{AS} =  -(N-2) \, \quad \mbox{\boldmath $\alpha$}_i  \mbox{\boldmath $\rho$}_{AS} = 1\,
  , \quad  i=1,\,  \ldots\,  N-2
\end{equation}
we observe that the vertex operators $ e^{i \mbox{\boldmath $\alpha$}_i\, 
\mbox{\boldmath $\sigma$}  }$ transform under the discrete shift 
symmetry 
$$
  \mbox{\boldmath $\sigma$}    \rightarrow  
 \mbox{\boldmath $\sigma$}  - \frac{2 \pi}{N-2} \mbox{\boldmath $\rho$}_{AS} \
 $$
 as
 \beqn
Z_{N-2}:  && e^{i   \mbox{\boldmath $\alpha$}_{2m}    \mbox{\boldmath 
$\sigma$}   }  
  \rightarrow   e^{i   \mbox{\boldmath $\alpha$}_{2m}    \mbox{\boldmath 
$\sigma$}   }, \qquad  e^{i   \mbox{\boldmath $\alpha$}_{2m+1}    \mbox{\boldmath 
$\sigma$}   }  
  \rightarrow   e^{i   \mbox{\boldmath $\alpha$}_{2m+1}    \mbox{\boldmath 
$\sigma$}   }  \,, \nonumber\\[3mm]
  &&  e^{i   \mbox{\boldmath $\alpha$}_{i}    \mbox{\boldmath 
$\sigma$}   }  
  \rightarrow  e^{- i \frac{2 \pi}{N-2}} \;   e^{i   \mbox{\boldmath $\alpha$}_{i}    
    \mbox{\boldmath 
$\sigma$}   }  \,,
\nonumber\\[3mm] 
&&
 i=1, \, \ldots \, 2m-1\,.
\eeqn
Hence, the monopole-induced interactions (\ref{Eq:orientivertex}) are invariant 
under $(Z_{2N-4})_A$ given in (\ref{Eq:AS}). The discrete shift symmetry allows mass terms for all dual photons at order $e^{-2S_0}$.  

In QCD(AS)*, there are novel topological excitations  as is the case in QCD(BF)*.  
The zero mode structure of monopole-instantons suggests that other than the magnetic bions common with SYM* theory, there are magnetic bions of a more exotic variety,
 \begin{eqnarray}
&&{\cal B}^1_i : \Big( \frac{4\pi}{g}(\mbox{\boldmath $\alpha$}_i\, - \mbox{\boldmath $\alpha$}_{i-1}) 
\,, \; 0 \Big)  :  \qquad \;\; \;\; c_1 e^{-2S_0} e^{i ( \mbox{\boldmath $\alpha$}_i \, - \mbox{\boldmath $\alpha$}_{i-1} \,) \mbox{\boldmath $\sigma$}} \,,  \nonumber\\[3mm]
&& {\cal B}^{12}_{i,i} :  \Big(  \frac{4\pi}{g}(\mbox{\boldmath $\alpha$}_i\,   - \mbox{\boldmath $\alpha$}_{2m-i})\,, 0 \Big)  : \qquad \;\;
  c_2 e^{-2S_0} e^{i ( \mbox{\boldmath $\alpha$}_i -  
  \mbox{\boldmath $\alpha$}_{2m-i} ) \mbox{\boldmath $\sigma$} 
  } \, , 
  \nonumber\\[3mm]
  && {\cal B}^{12}_{i,i-1} :  \Big( \frac{4\pi}{g}( \mbox{\boldmath $\alpha$}_i\,   - \mbox{\boldmath $\alpha$}_{2m-i+1})\,, 0 \Big)  : \quad 
  c_2 e^{-2S_0} e^{i ( \mbox{\boldmath $\alpha$}_i -  
  \mbox{\boldmath $\alpha$}_{2m-i+1} ) \mbox{\boldmath $\sigma$} 
  } \, , 
  \nonumber\\[3mm]
&& {\cal B}^{12}_{i, i+1} :  \Big(  \frac{4\pi}{g}(\mbox{\boldmath $\alpha$}_i\,   
- \mbox{\boldmath $\alpha$}_{2m-i-1})\,, 0 \Big)  : \quad 
  c_2 e^{-2S_0} e^{i ( \mbox{\boldmath $\alpha$}_i -  
  \mbox{\boldmath $\alpha$}_{2m-i-1} ) \mbox{\boldmath $\sigma$} 
  } \, .
\end{eqnarray}
Here in the first line summation runs over $ i=1,\,\ldots,  \, 2m-1$
while in the second, third and fourth lines over $i= 1,\, \ldots,\, m-1$. 
The pairing of the constituent  monopoles follows from the structure of the fermion zero modes. 
The  magnetic bion   ${\cal B}^1_i$ is held together due to the attractive fermionic pair exchanges which overcomes 
the Coulomb repulsion between its constituents. The constituents of the latter  bions 
 ${\cal B}^{12}_{i,i}$ and  ${\cal B}^{12}_{i,i\pm1}$ 
  do not interact via the Coulomb law, rather
they experience just the fermion pair exchange. Consequently, the combined effect of the magnetic bions (which is order $e^{-2S_0}$),
 \begin{equation}
 V_{\rm bion} (  \mbox{\boldmath $\sigma$} )   = m_W^3 g^{-6}  \left[ \sum_{i=1}^{2m-1} {\cal B}^1_i  + 
 \sum_{i=1}^{m-1} ( {\cal B}^{12}_{i,i} +  {\cal B}^{12}_{i,i+1} +  {\cal B}^{12}_{i,i-1} ) \right] + {\rm H.c.}
  \end{equation}
 and two monopole-instantons  
${\mathcal M}_{2m},  {\mathcal M}_{2m+1}$ gives rise to the bosonic potential which renders all $N-1$ dual photons massive, which, in turn, leads to string
(domain line) formation.   
Assembling perturbative and nonperturbative effects we get
\begin{eqnarray}
&&L^{\rm QCD(AS)^*} = \frac{g_3^2}{32 \pi^2}  (\partial  \mbox{\boldmath $\sigma$})^2 +
V_{\rm bion} (  \mbox{\boldmath $\sigma$} )  +    \;  \sum_{i=2m}^{2m+1}    \; ( {\cal M}_{i }  + \overline  {\cal M}_{i } ) 
\nonumber\\[3mm]
&& 
 + \frac{2}{g_3^2} \sum_{i=1}^{m}     \bar \Psi_i \gamma_{\mu} \Big( i \partial_{\mu} + 
 (  \mbox{\boldmath $H$}_{ii} + \mbox{\boldmath $H$}_{N-i,N-i} ) 
     \mbox{\boldmath $A$}_{\mu} 
 \Big)  
      \Psi_i  
       +  \;  \sum_{i=1}^{2m-1}    \; ( {\cal M}_{i }  + \overline  {\cal M}_{i } ) \,.
       \nonumber\\
\label{Eq:dQCD2} 
\end{eqnarray} 

In QCD(F/BF)* we had both electric couplings and monopole and bion-induced
magnetic interactions. By the same token in QCD(AS)* interactions of the
electric and magnetic type are present. (This is unlike what we have in SYM* theory.)
The monopole and bion-induced effects are dominant.
  
In the effective low-energy theory (\ref{Eq:dQCD2}),  the $(Z_{2N-4})_A$ chiral symmetry 
is entangled with the shift symmetry of the dual photon. Examination of the 
bosonic potential in QCD(AS)*  reveals $N-2$ gauge inequivalent isolated vacua located at
 \begin{equation}
 \mbox{\boldmath $\sigma$}
 = \left\{ 0, \, \frac{2 \pi}{N-2}, \, \frac{4 \pi}{N-2}, \,\ldots, \, \frac{2 (N-3)\pi}{N-2} \right\} 
 \mbox{\boldmath $\rho$}_{AS}\,.
 \end{equation}
As usual, we label these $N-2$ vacuum states by  $|\Omega_k\rangle$, $(k=1, \,\ldots \, ,N-2$).  Choosing a vacuum we spontaneously break the $Z_{N-2}$ symmetry. 
  
The chiral condensate  in the  vacuum $|\Omega_k\rangle $ can be calculated along the 
same lines as in QCD(BF)*, 
   \begin{eqnarray}
\langle \Omega_k | \tr \bar \Psi   \Psi  | \Omega_k \rangle    
=2(N-2) \left\{ \begin{array}{ll}
\Lambda^3    
(\Lambda L)^{4/3N}  \,, &  \; \Lambda  L  \ll1\,, \\[3mm] 
\Lambda^3  \,,    &  \; \Lambda  L  \gsim 1\,, \\ 
\end{array}
                                                        \!\right\}
                                                          \cos\!\left({\frac{2 \pi k}{N-2}}  \right),
\nonumber\\
\end{eqnarray}
where there is a weak $L$ dependence at small $L$. This follows from the 
$\O(1/N)$ difference in $b_0$, the first $\beta$-function coefficient of QCD(AS) and SYM
theories divided by $N$. In QCD(AS) 
$$
b_0=  3+  \frac{4}{3N}\,.
$$

{\bf Remark on the Callias and Atiyah--Singer index theorems:}
On $R_4$, the global aspect of the chiral anomaly is 
expressed by the Atiyah--Singer index theorem.  BPST instanton is associated with $2 h$ fermionic zero modes, where  
$2h=\{ 2, 2N, 2N, 2N-4, 2N+4 \} $ for QCD(F/adj/BF/AS/S), respectively.  
In QCD(${\mathcal R}$)* at small $r(S_1)$, due to the gauge symmetry breaking, the four-dimensional instanton splits into $N$ monopoles. 
In the small $r(S_1)$ (weak coupling) regime, the instanton should be viewed as a composite object, with the magnetic and topological charges as
in Eq.~(\ref{fracins}),
built of $N$ types of elementary monopoles with charges $\frac{4 \pi}{g} (\mbox{\boldmath $\alpha$}_1, \mbox{\boldmath $\alpha$}_2,\,  \ldots ,\, 
\mbox{\boldmath $\alpha$}_N)$.  The $2h$ fermion zero modes split into groups which associate themselves
 with the above $N$ monopoles as follows:
\begin{eqnarray}
&&
{\rm QCD(F)}: 2 \qquad \quad\rightarrow \{2, 0, \ldots, 0,0,0\}\,  , 
\nonumber\\[2mm] 
&&
{\rm SYM }: 2N  \qquad\quad \; \; \rightarrow \{2, 2, \ldots, 2, 2,2\}\, ,
  \nonumber\\[2mm] 
&&
{\rm QCD(BF)}: 2N \quad \;\;  \rightarrow \{2, 2, \ldots, 2,2,2\} \, ,
  \nonumber\\[2mm] 
&&
{\rm QCD(AS)}: 2N-4  \rightarrow \{2, 2, \ldots, 2, 0, 0\} \, ,
  \nonumber\\[2mm] 
&&
{\rm QCD(S)}: 2N +4  \;\;\;  \rightarrow \{2, 2, \ldots, 2, 4,4\}  \,.
\end{eqnarray}
The numbers on the right-hand side are the Callias indices for the corresponding monopoles.  Strictly  speaking, the Callias index theorem is formulated for 
the Yang--Mills + noncompact adjoint Higgs system 
on $\R^3$ \cite{Callias:1977kg}. Its generalization to $\R^3 \times S^1$ is 
carried out  by Nye and Singer, \cite{Nye:2000eg}. 
To study the index theorems we need to find the kernels 
of the Dirac operators $\Dslash$ and  $\Dslash^{\dagger}$ in the background of the appropriate topological excitation. 
The kernel is the set of zero eigenstates of the Dirac operator. 
The difference of the dimensions of the kernels gives the number of zero mode attached to a given topological excitation. Thus, we observe the following 
relation between the Atiyah--Singer index ${\cal I}_{\rm inst}$ and the Callias
index ${\cal I}_{\mbox{\boldmath $\alpha$}_i }$, 
\beq
{\mathcal I}_{\rm inst}=\sum_{\mbox{\boldmath $\alpha$}_i \in 
\Delta_{\rm aff}^{0}}   {\mathcal I}_{\rm \mbox{\boldmath $\alpha$}_i}\, , 
\eeq
or
\beq
\rule{0mm}{6mm}
\dim \ker \Dslash_{\rm inst} - \dim \ker \Dslash^{\dagger}_{\rm inst} = 
\sum_{\mbox{\boldmath $\alpha$}_i \in 
\Delta_{\rm aff}^{0}} \left( \dim \ker \Dslash_{ \mbox{\boldmath $\alpha$}_i } - \dim \ker \Dslash^{\dagger}_{ \mbox{\boldmath $\alpha$}_i} \right) \,.
\eeq

\section{\boldmath{$\theta$} dependence}
\label{s6}

There is one more interesting aspect of the theory which has not yet been discussed,
namely, {$\theta$} dependence. It is well-known that
in pure Yang--Mills theory on $R_4$ physical quantities, e.g. string tensions,
do depend on $\theta$, and physical periodicity in $\theta$ is $2\pi$.
Introduction of one massless quark in representation ${\mathcal R}$
eliminates {$\theta$} dependence of physical quantities
since one can eliminate the $\theta$ term
through an appropriate chiral rotation of the fermion field, as a result of the chiral anomaly.
This does not mean that various order parameters, e.g. the bifermion condensate,
are $\theta$ independent. If a small fermion mass term is added,
physical quantities acquire {$\theta$} dependence; all {$\theta$}-dependent
effects are proportional to the fermion mass $m$.

Let us ask ourselves what happens on $R_3\times S_1$, in deformed
theories. At first, let us consider pure Yang--Mills, assuming that
$\theta\neq 0$. Then the instanton-monopole induced vertices at  level $e^{-S_0}$
are
\beq
{\mathcal L} = e^{-S_0}\sum_{j=1}^N 
\,\mu_j\,
e^{i \, \mbox{\boldmath $\alpha$}_j \mbox{\boldmath $\sigma$}+i\theta/N}
+{\rm H.c.}\,.
\label{thone1}
\eeq
By globally shifting
\begin{equation}
 \mbox{\boldmath $\sigma$} \rightarrow \mbox{\boldmath $\sigma$} - \frac{\theta}{N} 
 \mbox{\boldmath $\rho$} 
 \label{shift}
 \end{equation}
 where $ \mbox{\boldmath $\rho$}$ is the Weyl vector, and using the identities ($\ref{iden}$), we can rewrite the instanton-monopole vertices in  the form 
\beq
{\mathcal L} = e^{-S_0}\sum_{j=1}^{N-1}  
\,\mu_j\,
e^{i \, \mbox{\boldmath $\alpha$}_j \mbox{\boldmath $\sigma$}}
+  \mu_N e^{-S_0} e^{i \, \mbox{\boldmath $\alpha$}_N \mbox{\boldmath $\sigma$} + i \theta} 
+{\rm H.c.}\,,
\label{thone2}
\eeq
where the  $2 \pi$ periodicity is more transparent.   In both Eqs.~(\ref{thone1}) and  (\ref{thone2}) the vacuum angle dependence is explicit.

Introducing one fundamental fermion, and localizing the fermionic zero mode into the monopole 
with charge $\alpha_N$ without loss of generality, 
 we get, instead of (\ref{thone1}) and  (\ref{thone2}) 
\beqn
{\mathcal L}
&=&
 \tilde \mu_{N}\, e^{-S_0} e^{i \, \mbox{\boldmath $\alpha$}_{N} \mbox{\boldmath $\sigma$}
 +i\theta/N } \,\lambda \psi + e^{-S_0}\sum_{j=1}^{N-1} 
\,\mu_j\,
e^{i \, \mbox{\boldmath $\alpha$}_j \mbox{\boldmath $\sigma$}+i\theta/N}
+{\rm H.c.}\,.
\nonumber\\[3mm]
&=&
 \tilde \mu_N\, e^{-S_0} e^{i \, \mbox{\boldmath $\alpha$}_{N} \mbox{\boldmath $\sigma$}
 +i\theta  } \,\lambda \psi + e^{-S_0}\sum_{j=1}^{N-1} 
\,\mu_j\,
e^{i \, \mbox{\boldmath $\alpha$}_j \mbox{\boldmath $\sigma$} }
+{\rm H.c.}\,.
\eeqn
where we used  (\ref{shift}) in passing to the second step.
It is clear in the latter form that 
the  $\theta$ dependence can be completely absorbed
in the fermion fields, 
\beq
\left\{ \psi\,,\,\lambda \right\} \to \left\{ \psi e^{-i\theta/2} \,,\,\lambda e^{-i\theta/2}\right\}\,.
\label{redef}
\eeq
 If the fermion mass term  $m \psi\lambda$ is added,
the $\theta$ dependence can no longer be absorbed
in the definition of the fermion field. Performing (\ref{redef})
we change the phase of the mass parameter. Correspondingly,
one can expect physical $\theta$ dependent effects
proportional to $m$, such as the vacuum energy density 
\beq
{\cal E} (\theta) \sim m \langle \bar \Psi \Psi \rangle \;  
 \cos\theta \; , 
 \eeq
in parallel with
the behavior of the undeformed theory on $R_4$. 

Analysis of the $\theta$ dependence in QCD(BF)* is even easier technically.  
The magnetic bion vertices have no $\theta$ dependence because 
each of them represent the product of a monopole and antimonopole vertex in 
which the $\theta$ dependence cancels. Moreover, the monopole-induced vertices are 
   \beqn
   && \Delta L^{\rm QCD(BF)^*}=
     e^{ -S_{0}} 
    \sum_{\mbox{\boldmath $\alpha$}_{i} \in \Delta_{\rm aff}^{0}}
    \Big(  \left(e^{i \mbox{\boldmath $\alpha$}_i\, \mbox{\boldmath $\sigma$}_1
     + i \theta/N } + 
     e^{i \mbox{\boldmath $\alpha$}_i\, \mbox{\boldmath $\sigma$}_2
       + i \theta/N }\right)   
    \nonumber\\[3mm]
    &&\times
    ( \lambda_i  \psi_i +  \lambda_{i+1}  \psi_{i+1}    )
      +  {\rm H.c.}
                  \Big)\,.
     \eeqn
The $\theta$ dependence can be readily absorbed 
in the fermion fields with the following redefinition:
\beq
\left\{ \psi_i\,,\,\lambda_i \right\} \to  
e^{-i\theta/(2N)} \
\left\{ \psi_i \,,\,\lambda_i \right\}\,.
\label{redef2}
\eeq
If we introduce very small mass terms for the fermion fields,
$m \ll \Lambda (\Lambda L)$, then it is obvious that the $\theta$ dependence reappears in the vacuum energy density,
\beq
{\mathcal E} (\theta) =  \min_k {\cal E}_k (\theta) \equiv  \min_k \left[ m \Lambda^3 \cos \left( \frac{\theta}{N}  + \frac{2 \pi k}{N} \right) \right], 
\quad k=1,\, \ldots, \, N\,.
\eeq
Turning on a nonvanishing mass term lifts the $N$-fold  degeneracy of the  vacua $|\Omega_k \rangle$. The vacuum labeled by the integer  $k$ turns into a state with energy   ${\cal E}_k (\theta)$. Each one of the 
 $N$ branches  is $2 \pi N$ periodic in $\theta$. Consequently, the vacuum energy density 
is physically $2 \pi$ periodic,
$$
{\mathcal E}_{\rm vac} (\theta + 2\pi)=  {\mathcal E}_{\rm vac} (\theta) 
\,.
$$ 
This is precisely the expected behavior of undeformed QCD(BF) on $R_4$.

In the case of QCD(AS)* the overall
picture emerging from our analysis is quite similar
(albeit there are some minor differences subleading in $1/N$)
and also matches the known $\theta$ dependence of QCD(AS) on $R_4$. 

\section{Remarks on planar equivalence} 
\label{plan}

Similarity of the dynamical aspects of QCD(BF/AS/S)* 
(with fermions in the two-index representation) and $\N=1$ SYM* theory
is evident.  Given that they are quantum theories with distinct matter content and distinct microscopic symmetries, this similarity is remarkable.
We explicitly showed that in the small $r(S_1)$ regime, QCD(BF/AS/S)* 
confine through the magnetic bion mechanism in the same way as $\N=1$ SYM* theory.  
Moreover, spontaneous breaking of the discrete chiral symmetries 
is similar in these two cases too. The bifermion condensate is saturated by 
a monopole-instanton with appropriate fermion zero mode structure.  
The calculated mass gaps are quite alike in both cases.
Clearly, our analysis makes it manifest that solvability of $\N=1$  SYM*
theory at weak coupling  is due to the unbroken center symmetry. 
Supersymmetry is secondary in this regime. 

In fact, an intimate relation between SYM theory and its orientifold-orbifold
daughters exists not only at small $r(S_1)$ but also in the decompactification limit
of large $r(S_1)$. If the number of colors $N\to \infty$,
there is a well defined equivalence between $\N=1$ SYM and  QCD(BF/AS/S) 
which goes under the name of planar equivalence 
\cite{Armoni:2003gp, Armoni:2004uu, Armoni:2004ub, Kovtun:2003hr, Kovtun:2004bz}. 
The necessary conditions for planar equivalence to be valid nonperturbatively
are (i) interchange $(Z_2)_I$ symmetry is unbroken in QCD(BF), ii) $C$ conjugation symmetry is unbroken in  QCD(AS/S). It is generally believed
that these conditions are met \cite{Shifman:2007kt}.

The large $N$  equivalence is a useful tool to translate nonperturbative 
data of SYM theory to its daughters (and vice versa) on $R_4$. Planar equivalence
is valid also on $R_3 \times S_1$.  The equivalence establishes
an isomorphism on a subspace of the Hilbert space of these theories. 
Let us grade the Hilbert space of SYM theory with respect to
$(-1)^F$ where $F$ is the fermion number, as 
\beq
{\cal H}^{\rm SYM } =  {\cal H}^{\rm SYM +} \oplus  {\cal H}^{\rm SYM -}\,.
\eeq
Similarly, the Hilbert spaces of QCD(BF) and QCD(AS/S)
can be graded respect to the $1\leftrightarrow 2$ interchange  symmetry
in the first case and 
charge conjugation in the second.
Planar equivalence is an isomorphism between the even subspaces of the Hilbert spaces
\begin{equation}
  {\cal H}^{\rm SYM +}  \equiv   {\cal H}^{\rm {QCD(BF)}+} \equiv   
  {\cal H}^{\rm {QCD(AS)} +} \,.
\end{equation}
(The full Hilbert spaces are by no means isomorphic.)

If one performs periodic compactifications\,\footnote{In thermal compactification, 
only the center symmetry breaks 
spontaneously; the interchange symmetry  and $C$ 
invariance remain unbroken \cite{Unsal:2006pj}. Thus, planar equivalence 
for orbifold and orientifold daughters remains valid in the high temperature deconfined phase.} of QCD(BF/AS/S) on $R_3 \times S_1$, with small $r(S_1)$, the $1\leftrightarrow 2$ interchange  symmetry of QCD(BF)* and $C$ invariance of QCD(AS/S)* do break spontaneously, along with 
the spatial center symmetry \cite{Tong:2002vp, Unsal:2006pj}. (For related lattice studies 
showing the breaking and restoration of $C$ see \cite{DeGrand:2006qb, Lucini:2007as}.) 

Certain  order parameters which probe the interchange symmetry and  $C$ invariance are topologically nontrivial \cite{Unsal:2007fb}, e.g.
\beq
 \tr (U_1^k)  - \tr (U_2^k), \,\,\,\, {\rm QCD(BF)^*} \,\,\,\,  \mbox{and}\,\,\,\,
\tr (U^k)  - \tr ( U^{*\,k}) \,\,\,\,{\rm QCD(AS)^*} \,.
\label{order}
\eeq
These operators are charged under the center symmetry and odd under $(Z_2)_I$ and  $C$. 
In QCD(BF/AS/S)* stabilization of the center symmetry automatically implies vanishing of 
the expectation values of the order parameters (\ref{order}). 

There are also order parameters which are neutral under the center symmetry, yet charged under $(Z_2)_I$ and $C$. For example, the odd combination of  
the Wilson loops $W_1(C) -W_2(C)$ or 
$\tr F_1^2 - \tr F_2^2$ in QCD(BF)* and $W(C) -W^*(C) $ in QCD(AS)* are of this type. The unbroken center symmetry does not restrict the expectation value of such operators. Our dynamical analysis in Sects. (\ref{s4}) and (\ref{s5}) shows that spontaneous breaking of $(Z_2)_I$ and $C$ symmetry definitely
does not take place at small $r(S_1)$. Arguments why this must be the case also on
$R_4$ are summarized in Ref.~\cite{Shifman:2007kt}. 

\section{Conclusions and prospects: \\ Abelian vs. 
non-Abelian confinement}
\label{s7}

The aspects of QCD* theories that we studied are valid in the 
limit $L \Lambda   \ll 1$, where the weak coupling regime sets in.  
We presented arguments that one-flavor QCD($\mathcal R$)* theories 
are continuously connected to one-flavor QCD($\mathcal R$) on $R_4$.   
We demonstrated, through an explicit calculation
at small $r(S_1)$, existence of  the mass gap, linear confinement, 
and discrete $\chi$SB. These are indeed the most salient features of QCD-like
theories on $R_4$.  

In the small $r(S_1)$ domain, the QCD* theories are characterized
by the fact that the gauge symmetry is Higgsed down to
a maximal Abelian subgroup $U(1)^{N-1}$. 
Thus, at small $r(S_1)$ we deal with Abelian confinement, while it
is expected to give place to non-Abelian confinement in the decompactification
limit. 

What happens as we increase
$ L \Lambda $ gradually, all the way to $L\to\infty$? At a scale of the order 
$ L \Lambda  \sim 1$, we loose the separation of scale between the $W$-bosons  
and the nonperturbatively gapped photons. Thus, our effective low-energy 
description (which includes only light bosonic and fermionic degrees of freedom)
ceases to be valid. At and above $\Lambda\sim 1/ L $ the theory is strongly coupled in the IR,  and the full non-Abelian gauge group is operative. Thus, the confinement mechanism in this regime must be non-Abelian. 

This situation is completely analogous to the Seiberg--Witten solution \cite{Seiberg:1994rs}
of four-dimensional $\N=2$ SYM  theory exhibiting mass gap and linear confinement upon a $\mu$ deformation breaking
$\N=2$ down to $\N=1$.   If $\mu/\Lambda \ll 1$, the Seiberg--Witten theory 
in the IR is in the regime of broken gauge symmetry, i.e.
SU$(N)\to {\rm U}(1)^{N-1}$, where it is solvable. For $\mu/\Lambda \gsim  1$, one looses the separation of scales between the $W$ bosons and nonperturbatively gapped photons. 
The full gauge symmetry is restored. In this regime, the low-energy theory approaches 
pure $\N=1$ SYM theory. The confining strings must be non-Abelian. Currently no 
controllable analytical approaches allowing one to continue the
Seiberg--Witten solution to the domain $\mu/\Lambda \gg  1$ are known, 
and yet there are good reasons to believe that this continuation is smooth.

Conceptually the relation between $\mu$-deformed $\N=2$  and $\N=1$ SYM 
theories on $R_4$ is parallel to that between one-flavor QCD* on $R_3 \times S_1$ and    
QCD on $R_4$. Both theories realize confinement via the following pattern
   \begin{eqnarray}
{\rm SU}(N) \; \; \; \stackrel{\rm  Higgsing}{\longrightarrow} \; \; \;  [{\rm U}(1)]^{N-1}
 \; \; \; \stackrel{\rm nonperturbative}{\longrightarrow} {\rm no\,\, massless  \,\, modes}  \,.
\label{pattern1}
  \end{eqnarray} 
Existence of an intermediate Abelian gauge theory in the IR is the key to  analytical calculability in both cases. 
 
In both cases by tuning the relevant parameter, $\mu/\Lambda$ or $L\Lambda$,  respectively, 
from small to large values, we can remove the intermediate step of ``Abelianization."
In this paper we presented a number of arguments in favor
of no phase transitions separating the Abelian and non-Abelian confinement regimes. It is desirable to develop a special technique allowing one to perform   ``integrating in" of the 
$W$ bosons (and their partners) gradually.  If this task can be achieved this could provide a direct route to QCD and  QCD-like theories on $R_4$.

If we are right and the transition from QCD* to QCD-like theories is smooth,
this smoothness could explain a long-standing puzzle.
The point is that a rather vaguely defined method which goes under the name of the
maximal Abelian projection seems to give sensible results in the lattice calculations.
The reason might be the proximity of the Abelian confinement regime
we discussed in the body of this paper.

The status of QCD-like theories with massless or very light fermions with  exact 
or approximate chiral symmetry significantly improved in the recent 
years~\cite{Kaplan:1992bt,Narayanan:1994gw}. It is highly
desirable to implement QCD* theories on lattices, and then carry out an in-depth
study of the transition from Abelian to non-Abelian confinement. 
 
\section*{Acknowledgments}
M.\"U. thanks E. Silverstein for discussions on double trace deformations, and 
D. T. Son for  discussions on Polyakov's model.  M.S. is grateful to A. Yung for endless discussions of Abelian vs. non-Abelian confinement. We would like
to thank Adi Armoni for stimulating questions and fruitful correspondence.
The work of M.S. is supported in part by DOE grant DE-FG02-94ER408. 
The work of  M.\"U. is supported by the U.S.\ Department of Energy Grant DE-AC02-76SF00515.

\newpage

\section*{Appendix: Center stabilization} 
\label{app}

\addcontentsline{toc}{section}{Appendix: Center stabilization}

\renewcommand{\theequation}{A.\arabic{equation}}
\setcounter{equation}{0}

Let $U({\bf x})$ be the path-ordered 
holonomy of the Wilson line wrapping $S_1$ at the point 
${\bf x} \in R_3$.  It is known that for complex representation fermions (F/AS/S/BF), the center symmetry is broken down at sufficiently small $r(S^1)$ regardless of the spin connections of fermions. For adjoint fermions with periodic spin connection, 
the spatial center symmetry is not broken at small $r(S_1)$, whereas for antiperiodic (thermal) boundary conditions the temporal center symmetry is broken at sufficiently high temperatures. 

An easy way to see this is to evaluate the one-loop Coleman--Weinberg effective potential induced by quantum fluctuations  by using the background field  method  
(e.g.  \cite{Gross:1980br,Unsal:2006pj}). 
The minimum of the {\em classical} action is achieved at the vanishing value of 
the gauge field strength, and constant but arbitrary values of the $U(\bf x)$. 
Quantum corrections lift the degeneracy.

One can evaluate the one loop-potentials  for one flavor QCD-like theories. In the gauge in which the Polyakov line is represented by a constant and diagonal
matrix one obtains\,\footnote{In the multiflavor generalization (with $N_f$ fermions) one must replace 
$a_n \rightarrow a_n N_f.$}
\beq
V_{\rm eff}[U]=
 \frac{2}{\pi^2 L^4}  \sum_{n=1}^{\infty} \frac{1}{n^4} \, T_n\,,
\eeq
where
\beqn
  &&  T_n=          -  |\tr \, U^n|^2   +   a_n  (\tr\, U^n + \tr \,U^{*\, n}) \,, \qquad {\rm (F) } \,,
\nonumber\\[3mm]
  &&  T_n=     
             (-1 +  a_n) |\tr\,  U^n|^2 \,, \qquad \qquad\qquad \qquad{\rm  (adj) } \,,
 \eeqn
 \beqn
 &&  T_n=           \frac{1}{2} (-1 +  a_n) |\tr \,U_1^n + \tr \,U_2^n|^2  
 \nonumber\\[3mm]
&& +       \frac{1}{2} 
              (-1 -   a_n) |\tr \,U_1^n - \tr \,U_2^n|^2 \,,\qquad\,\,\,\,\,
              \qquad {\rm  (BF) } \,,
\eeqn
\beqn
&&    T_n=             \frac{1}{4}  (-1 +  a_n)  |\tr \, U^n + \tr \,U^{*n}|^2   +    
                 \frac{1}{4}    (-1 -  a_n)  |\tr \,U^n - \tr \,U^{*n}|^2
 \nonumber\\[3mm]
     &&              \mp  \frac{1}{2} a_n  \left( \tr \, U^{2n} + \tr \,U^{*2n}\right) ,  \qquad 
           \qquad \qquad\qquad \qquad \qquad  {\rm  (AS/S) }\,.  
\label{eq:potential}
\eeqn
Here $a_n$ are prefactors which depend on the fermion boundary conditions,
\begin{equation}
a_n= \left\{ \begin{array}{cl}
                      (-1)^n  & {\rm for}\;  {\cal S}^{-}\,,\nonumber\\[3mm]
                       1  & {\rm for} \; {\cal S}^{+}\,.
                       \end{array}
                       \right.                      
\end{equation}
Note that
\begin{eqnarray}
 && 
 C\,\left(  \tr\, U^{n} \pm \tr\, (U^{ *})^n \right) = \pm \left( \tr\, U^{n} \pm \tr\, (U^{ *})^n\right)  \,,\nonumber\\[3mm]
&&
 {\mathcal I}\, \left( \tr \, U_{1}^{n} \pm \tr\,  (U_{2})^n\right) = \pm \left( \tr \, U_{1}^{n} \pm \tr \, (U_{2})^n\right).
 \end{eqnarray}

The minimum 
of the effective potential presented above is located at
\begin{eqnarray}
&&U \sim  {\rm Diag}(1, 1, \ldots\,, 1) 
\quad  {\rm all}\;   {\cal R} \; {\rm with} \;     { \cal S^-}  \;\;  {\rm and}  
\;  {\rm F/BF/AS/S}   
   \; {\rm with} \;   {\cal S^+} , 
\nonumber\\[3mm]
  && 
  U =  {\rm Diag} \left( 1,  e^{i  \frac{2\pi}{N}}, \ldots , 
 e^{i \frac{2\pi (N-1)} {N}}  \right)  
\quad  {\rm adj \;   with} \; \;  { \cal S^+}  \, .
\end{eqnarray} 
Thus,  the (spatial or temporal)  center symmetry is broken in all theories, except QCD(adj) with the periodic spin connection ${\cal S}^{+}$. 
In the cases of broken center symmetry the small and large radius physics on $S_1 \times R_3$ are separated by a phase transition. In all these cases   
the fermions essentially decouple from infrared physics,  and the theory  
at small $r(S_1)$  has not much in common with the theory at large $r(S_1)$. 
   
The center symmetry breaking is induced by destabilizing double trace operators 
such as e.g.
$- |\tr U|^2$ and their multiwinding counterparts. 
One can stabilize the center symmetry while respecting the underlying symmetries of 
the theories at hand by adding a stabilizing polynomial in the appropriate variable 
up to the winding number $[N/2]$ with judiciously chosen coefficients. This will overwhelm the one-loop effect, and make the center-symmetric point a 
stable vacuum in the small $r(S_1)$ regime.


\end{document}